\documentclass[a4paper,11pt,pdftex] {article}

\usepackage[latin1]{inputenc}
\usepackage{amsmath,amsfonts}
\usepackage{caption}
\usepackage{subcaption}

\usepackage{wasysym}
\usepackage{xspace}
\usepackage{bm}
\usepackage{color}
\usepackage{multirow}
\usepackage{url}
\usepackage[final]{pdfpages} 
\usepackage{hyperref}
\usepackage[normalem]{ulem}

\newcommand{\code}[1]{\small \texttt{#1}}

\newcommand{\joint}[1]{#1}
\newcommand{\snr}{\mathrm{SNR}}

\newcommand{\add}[1]{#1}
\newcommand{\suppress}[1]{}

\graphicspath{
  {fig/} 
  {fig/AUC/} 
  {fig/SNR/}
}

\usepackage{authblk}

\begin{document}

\title{Performance evaluation of DNA copy number segmentation methods}

\author[1]{Morgane Pierre-Jean\thanks{morgane.pierrejean@genopole.cnrs.fr}}
\author[2]{Guillem Rigaill\thanks{rigaill@univ-evry.fr}}
\author[1]{Pierre Neuvial\thanks{pierre.neuvial@genopole.cnrs.fr}}
\affil[1]{Laboratoire de Mathématiques et Modélisation d'Évry, Université d'Évry val d'Essonne, UMR CNRS 8071, USC INRA}
\affil[2]{Unité de Recherche en Génomique Végétale (URGV), Université d'Évry val d'Essonne, UMR INRA 1165 - CNRS 8114}
\renewcommand\Authands{ and }

\maketitle
\section*{Biographical notes}
\noindent Morgane Pierre-Jean is a PhD student in the Statistics and Genome team of the Laboratoire de Mathématiques et Modélisation d'Évry, University of Évry val d'Essonne / CNRS / INRA, France.  She received her MS Degree in statistics from the University of Rennes in 2012.\\

\noindent Guillem Rigaill is an Assistant Professor at the University of Évry val d'Essonne, working in the Bioinformatics for Predictive Genomics team of the URGV /  UMR INRA 1165 - CNRS 8114, France.\\

\noindent Pierre Neuvial is a CNRS researcher in the Statistics and Genome team of the Laboratoire de Mathématiques et Modélisation d'Évry, University of Évry val d'Essonne / UMR CNRS 8071 /  USC INRA, France.\\

\section*{Key Points}

\begin{itemize}
\item A number of methods are available for segmenting DNA copy number profiles in cancer studies.
\item A robust and reproducible comparison of such methods requires the definition of a framework for generating realistic copy number profiles, and a framework for assessing methods' performance.
\item A data generation framework based on resampling from real data makes it possible to compare different methods across a large number of different realistic scenarios.
\item The performance of segmentation methods is mainly driven by biological parameters such as the proportion of tumor cells in the sample and the proportion of heterozygous markers.
\item Using the open source and cross-platform {\code{R}} package {\code{jointseg}}, the present comparison study may be reproduced either on the data sets provided or on other data sets. 
\end{itemize}

\begin{abstract}
A number of bioinformatic or biostatistical methods are available for analyzing DNA copy number profiles measured from microarray or sequencing technologies.  In the absence of rich enough gold standard data sets, the performance of these methods is generally assessed using unrealistic simulation studies, or based on small real data analyses.  

In order to make an objective and reproducible performance assessment, we have designed and implemented a framework to generate realistic DNA copy number profiles of cancer samples with known truth.  These profiles are generated by resampling \suppress{real} \add{publicly available} SNP microarray data from genomic regions with known  copy-number state.  The original data have been extracted from dilutions series of tumor cell lines with matched blood samples at several concentrations.  Therefore, the signal-to-noise ratio of the generated profiles can be controlled through the (known) percentage of tumor cells in the sample.

This paper describes this framework and its application to a comparison study between methods for segmenting DNA copy number profiles from SNP microarrays.  This study indicates that no single method is uniformly better than all others.  It also helps identifying pros and cons of the compared methods as a function of biologically informative parameters, such as the fraction of tumor cells in the sample and the proportion of heterozygous markers.\\

This  comparison study may be reproduced using the open source and cross-platform {\code{R}} package {\code{jointseg}}, which implements the proposed data generation and evaluation framework: \url{http://r-forge.r-project.org/R/?group_id=1562}.
\end{abstract}

\paragraph{Keywords:}
  DNA copy number, segmentation, realistic data generation, performance evaluation.


\section{Background}
\label{sec:background}

Changes in DNA copy numbers are a hallmark of cancer cells \cite{hanahan11hallmarks}. Therefore, the accurate detection and interpretation of such changes are two important steps toward improved diagnosis and treatment.  The analysis of copy number profiles measured from high-throughput technologies such as array-comparative genomic hybridization (array-CGH), Single Nucleotide Polymorphism array (SNP array) or high-throughput DNA sequencing data raises a number of statistical and bioinformatic challenges.

Various methods have been proposed in the past decade for analyzing such data.  \suppress{Form} \add{From} a practitioner's point of view, it is quite difficult to find which method is best for a given scientific question.  
In fact, it is likely that the overall difficulty of the problem depends on the context (technology, type of cancer, percentage of tumor cells). It is also likely that certain methods are more appropriate for certain contexts. Therefore, it is important to take this context into account when evaluating a set of methods, in order to 1) get a sense of the overall difficulty of the problem when interpreting the results and 2) choose appropriate methods for this context.
\add{Typically, a practitioner chooses among available data analysis methods or calibrates their parameters using a trial and error approach.  A limitation of such an approach is that it is subjective, hardly reproducible and non quantitative.}

\add{The present work tackles this problem by proposing a reproducible framework for evaluating the performance of existing segmentation methods for identifying change-points from DNA copy number profiles from cancer patients.}
\suppress{The present work is motivated by the problem of comparing the performance of existing segmentation methods for identifying change-points from DNA copy number profiles from cancer patients.}  
As any performance evaluation strategy, addressing this question requires the definition of three objects:
\begin{enumerate}
\item data with known ``truth'';
\item methods to be compared;
\item criteria for performance assessment.
\end{enumerate}
In this paper, we propose such a definition and illustrate how it may be used to compare segmentation methods.  The main contributions of this work are
\begin{itemize}
\item a framework to generate realistic DNA copy-number profiles with known ``truth''.  This framework is generic and may be applied to any copy number data set;
\item a framework to address the question of which SNP array data segmentation method performs best, depending on biologically relevant parameters. 
\end{itemize}
These frameworks are implemented in the {\code{R}} package {\code{jointseg}}. 
The rest of this paper is organized as follows. We start by giving some background on DNA copy number segmentation (Section \ref{sec:copy-numb-segm}) and  describe our proposed data generation framework (Section \ref{sec:generating-data}). Then, we describe the pipeline we use for evaluating segmentation methods (Section~\ref{sec:segm-pipel}). Finally, the result of our comparison study on two data sets are reported in Section~\ref{sec:results}. 

\section{DNA copy number segmentation}
\label{sec:copy-numb-segm}

\subsection{DNA copy number data}
\label{sec:dna-copy-number}
\begin{figure}[!h]
  \centering
  \includegraphics[width=8cm]{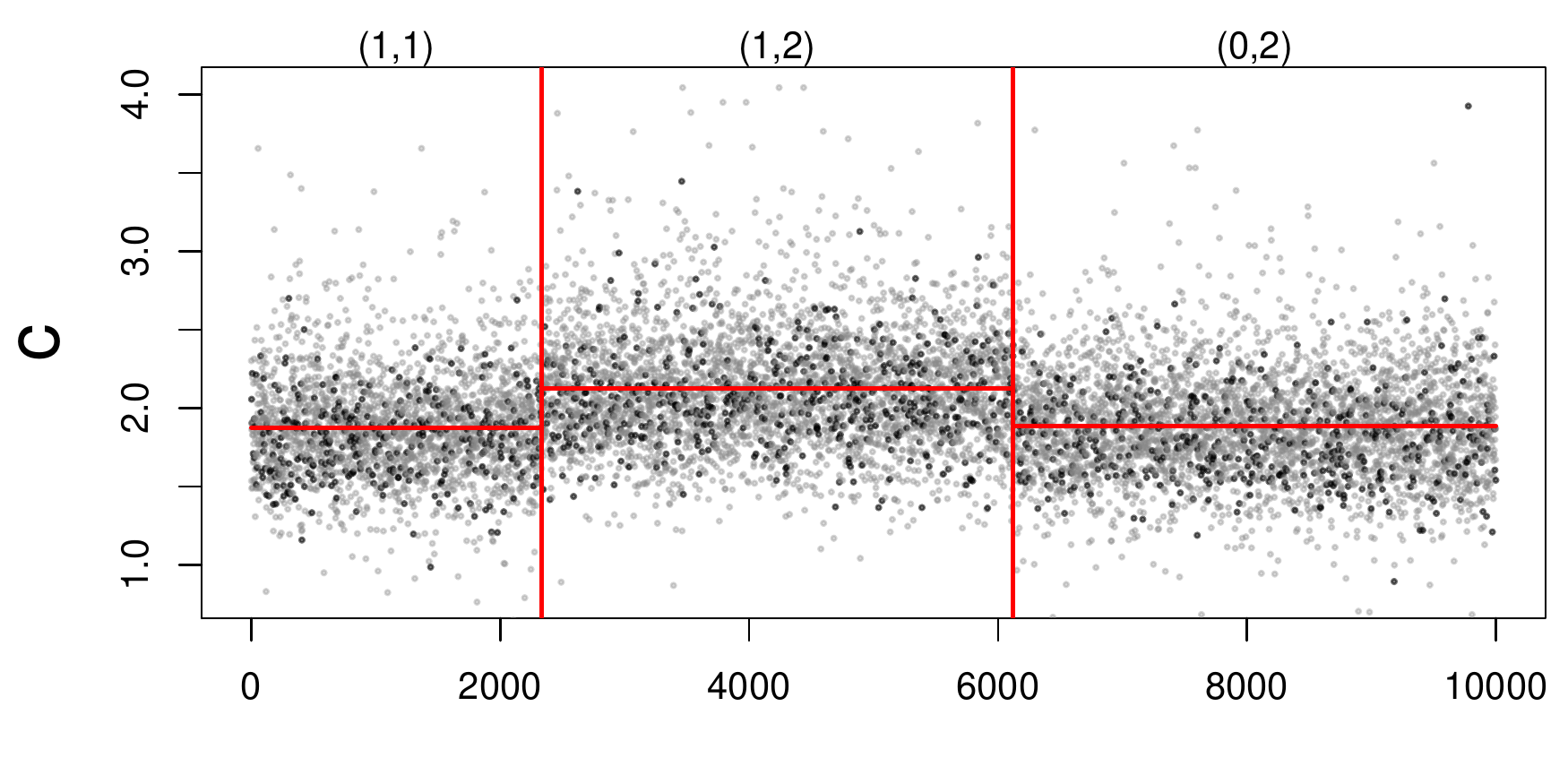}
  \includegraphics[width=8cm]{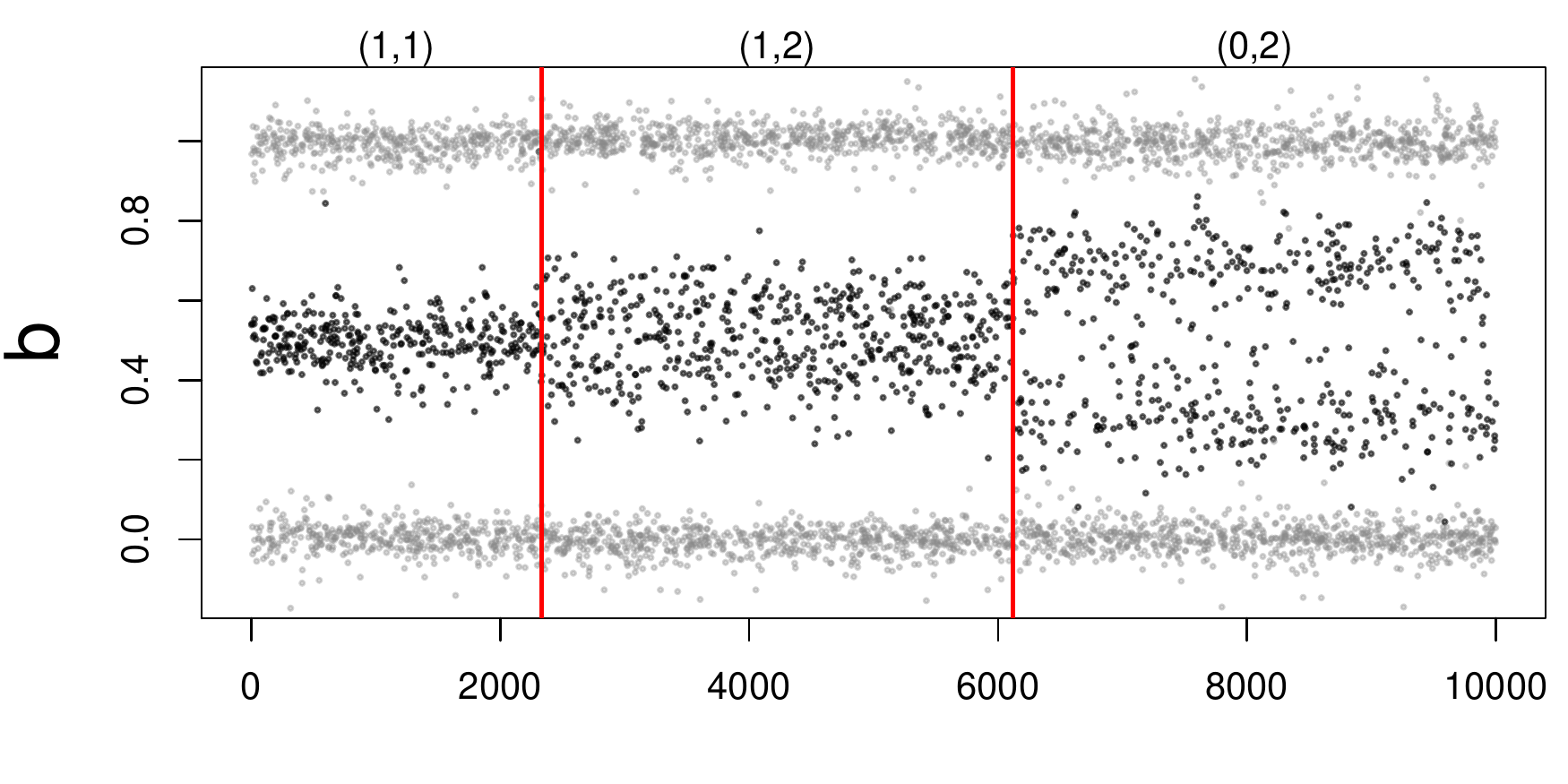}
\caption{Example SNP array data. Total copy numbers ($c$), allelic ratios ($b$) along 10,000 genomic loci. Red vertical lines represent change points, and red horizontal lines represent mean signal levels between two change points.  SNPs that are \suppress{homozygous} \add{heterozygous} in the germline are colored in black; all of of the other loci are colored in gray.}
\label{fig:copy-number-data-c-b}
\end{figure}

Normal cells have two copies of DNA, inherited from each biological parent of the individual. In tumor cells,  parts of a chromosome of various sizes (from kilobases to a chromosome arm) may be deleted, or copied several times.  As a result, DNA copy numbers in tumor cells are piecewise constant along the genome. Copy numbers can be measured using microarray or sequencing experiments.  For illustration, Figure~\ref{fig:copy-number-data-c-b} displays an example of copy number signals that may be obtained from SNP-array data.  Red vertical lines represent change points.  In this particular example, the first region [0-2200] is normal, the second one [2200-6100] is a region where one of the parental chromosomes has been duplicated, and the third one [6100-10000] is a region of uniparental disomy, that is, a region where one of the parental chromosomes has been duplicated and the other one deleted.  
The top panel represents estimates of the total copy number (denoted by $c$). The bottom panel represents estimates of allelic ratios (denoted by $b$).  We refer to  \cite{neuvial11statistical} for an explanation of how these estimates may be obtained.  In the normal region [0-2200], the total copy number is centered around two copies and allelic ratios have three modes centered at 0, 1/2 and 1.  These modes correspond to homozygous SNPs AA ($b=0$) and BB ($b=1$), and heterozygous SNPs AB ($b=1/2$).  \add{We note that in the second region where the tumor has 3 copies of DNA, the average observed signal is substantially below the true copy number.  This is due to the presence of normal cells in the ``tumor sample'', a phenomenon known as \emph{normal contamination} which shrinks the observed signals toward two copies of DNA.  The reader is referred to \cite{neuvial11statistical} for a more detailed explanation of this phenomenon and other sources of non-calibration in DNA copy number signals, such as the ploidy of the tumor.}
One important observation is that change points occur at the same position in both dimensions.  This is explained by the fact that a change in only one of the parental copy numbers is reflected in both $c$ and $b$.  Therefore, it makes sense to analyze both dimensions of the signal jointly in order to identify change points.

\begin{figure}
  \centering
  \includegraphics[width=8cm]{copyNumberData,c}
  \includegraphics[width=8cm]{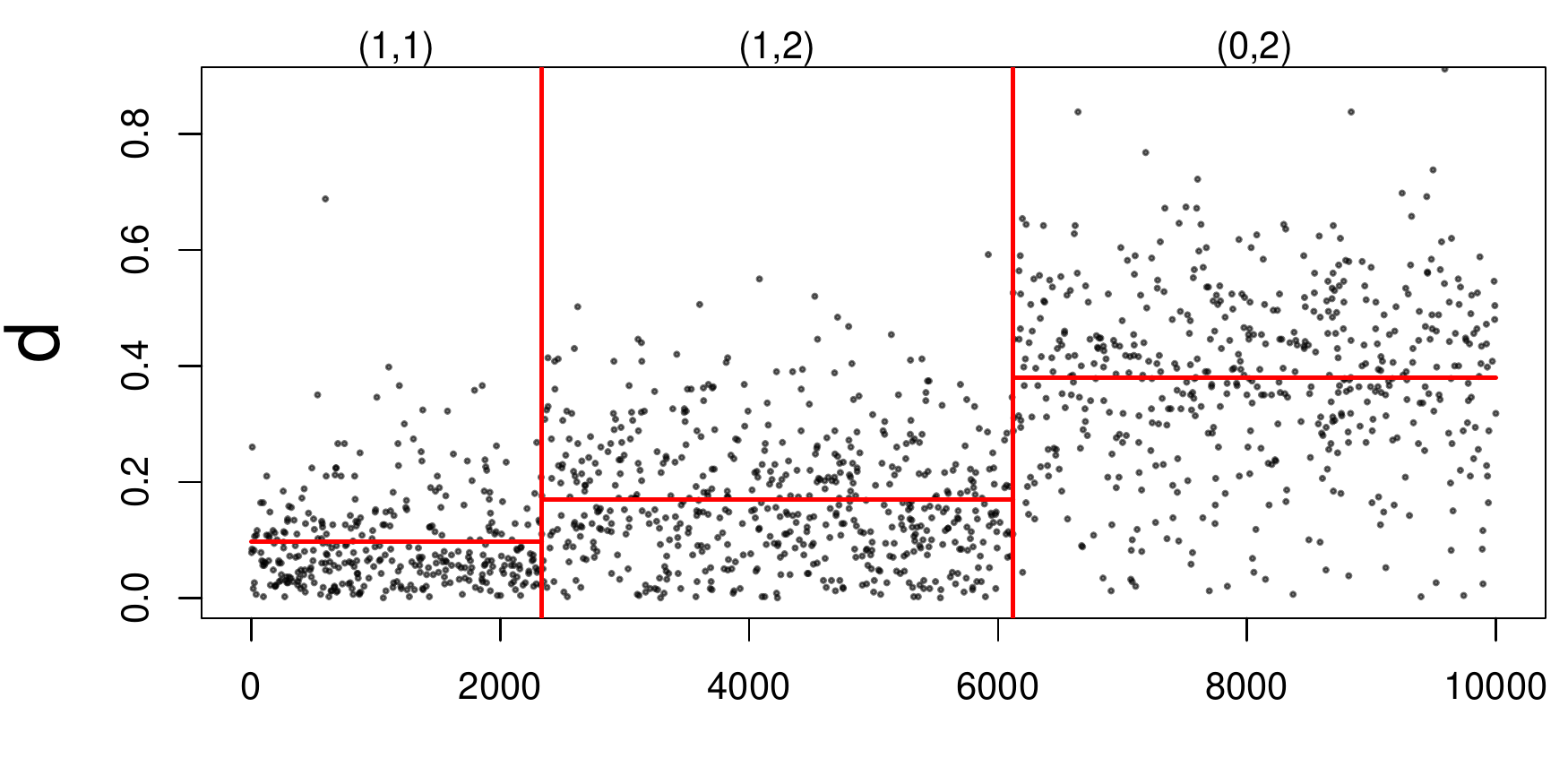}
\caption{Example SNP array data  along 10,000 genomic loci, after transformation of allelic ratios (b) into decrease in heterozygosity $(d)$, following~\cite{bengtsson10tumorboost,staaf08segmentation}. Red vertical lines represent change points, and red horizontal lines represent mean signal levels between two change points. SNPs that are \suppress{homozygous} \add{heterozygous} in the germline are colored in black; all of of the other loci are colored in gray.}
\label{fig:copy-number-data-c-d}
\end{figure}

In order to facilitate segmentation, allelic ratios ($b$) are generally transformed into unimodal signals, as originally proposed in~\cite{staaf08segmentation}.   This transformation is motivated by the fact that allelic ratios can be symmetrized (``folded'') and that SNPs that are homozygous in the germline (these SNPs are plotted in gray in Figure~\ref{fig:copy-number-data-c-b}) can be discarded as they do not carry any information about copy-number changes.  Following \cite{bengtsson10tumorboost}, we define the ``decrease in heterozygosity'' $d= 2\vert b-\frac{1}{2}\vert$ for SNPs that are heterozygous in the germline \add{(referred to as ``heterozygous SNPs'' in the remainder of the paper for short)}, which is essentially a rescaled version of the ``mirrored/folded BAF'' defined by ~\cite{staaf08segmentation}.  After this transformation, DNA copy numbers can be considered as a bivariate, piecewise-constant signal, as illustrated by Figure~\ref{fig:copy-number-data-c-d}.
\add{It should be emphasized at this stage that because the proportion of heterozygous markers among SNPs is generally of the order of 1/3 for a given sample, the number of informative markers is several times larger for $(c)$ than for $(d)$.  This feature of SNP array data has implications in terms of speed and performance of segmentation methods, which will be explained in detail later in the paper.}

\subsection{Typology of copy number segmentation methods}
\label{sec:typology-copy-number}

Many different methods have been proposed for the analysis of DNA copy number profiles.
Most of them may be classified into four categories: methods based on Hidden Markov Models (HMM),  multiple change-point methods, fused lasso-based methods and recursive segmentation methods.
\begin{enumerate}
\item HMM-based approaches rely on the idea that the recovered DNA copy
  number should be discrete and that these different levels can
  be modeled using a small number of HMM states. A typical example of such an HMM is the
  work of \cite{fridlyand04hidden}.  For the specific case of SNP array analysis in cancer samples, several dedicated HMM have been proposed~\cite{sun09integrated,greenman10picnic,chen11estimation}.
\item Multiple change-point methods assume that the observed signal is affected by abrupt changes and that between these breaks the signal should be homogenous  \cite{picard05a-statistical}.
\item Methods based on a fused lasso penalty rely on the idea that, in
  most cases, two successive measurements should have the same
  estimate. This is encoded by a $L_1$ penalty on successive differences. The recovered signal is
  guaranteed to be piecewise constant. A typical example of such a
  fused model is the work of \cite{tibshirani05sparsity}.  This class of methods can be viewed as solving a convex relaxation of the multiple change point problem. 
\item Recursive segmentation approaches rely on the intuitive idea that a
  segmentation can be recovered by recursively cutting the signal into
  two or more pieces. A typical example of such an recursive approach
  is the work of \cite{olshen04circular}.
\end{enumerate}
We refer to \cite{neuvial11statistical,zhang10dna-copy} for a more mathematical introduction to these methods.  Here, we only note that all of these methods assume that the signals are Gaussian. 
The above classification is by no means exhaustive (see for example \cite{hupe04analysis,ben-yaacov08fast}).

\section{Generating data with known ``truth''}
\label{sec:generating-data}

\subsection{Review of existing approaches}
A number of data generation mechanisms have been proposed in the context of performance evaluation of DNA copy number analysis in cancer samples, either in comparison studies~\cite{willenbrock05a-comparison,mosen-ansorena12comparison,lai05comparative,hocking13learning}, or in papers describing new analysis tools.  The generation of data with known ``truth'' can be done using either simulated or real data, both of which have \suppress{opposite} assets and drawbacks.

At first glance, simulated data are more appealing than real data because (i) ``truth'' is known with no ambiguity, (ii) the level of difficulty of the problem can be tuned as desired, and (iii) a large number of simulated data sets can be generated.  
As most DNA copy number segmentation methods rely on a Gaussian model (see Section~\ref{sec:copy-numb-segm}), their performance is usually assessed using Gaussian simulations~(see, for example, \cite{picard05a-statistical,zhang07modified}). 
While we do not question the usefulness of model assumptions for building statistical methods and for testing implementations, we believe that performance evaluation should as much as possible avoid \suppress{to rely}\add{relying on} on a particular model. 
A recent study which compared several approaches for segmenting univariate DNA copy number profiles using the multiple change point approach showed that the best performing methods on Gaussian simulations performed quite poorly on real data~\cite[Table 3]{rigaill13learning}.
In the remainder of this section, we briefly review some existing approaches that have tried to take the best of both the ``simulated data'' and the ``real data'' worlds:

\paragraph{An automatically annotated data set \cite{willenbrock05a-comparison}.}
The authors analyzed real data using one particular segmentation method to generate ``truth''.  They then used resampling to generate realistic copy-number profiles, where (Gaussian) noise was added in order to control the signal-to-noise ratio of the data set.  Two drawbacks of this approach are that  the notion of ``truth'' depends on the chosen segmentation method, and that the problem difficulty is not driven by biological considerations.

\paragraph{A dilution series \cite{staaf08segmentation}.}
In order to address the latter point, \cite{staaf08segmentation} have produced a dilution data set, where DNA from a lung cancer cell line is mixed with matched blood DNA from the same patient with varying (and known) mixture proportion (see description in Appendix~\ref{sec:affymetrix-data}).  Therefore, the fraction of tumor cells in the mixture controls the difficulty of the problem.   The ``truth'' is a panel of regions whose DNA copy number status in the cell line (normal, gain, hemizygous deletion, copy-neutral LOH \dots) is known.  This evaluation method has been accepted as a \textit{de facto} standard and has been used in several subsequent papers, including~\cite{chen11estimation,olshen11parent-specific,rancoita10integrated}.

An important drawback of this evaluation framework is that it focuses on a very limited number of regions (ten), which results in very little discrimination between most methods in realistic settings.  For example, four of the six methods compared in \cite{olshen11parent-specific} reach maximum sensitivity in all 10 regions for tumor cell fractions greater than 25\%.  In practice, samples with less than 50\% are rarely analyzed, in particular because the performance of most methods typically decreases severely when the fraction of tumor cells is less than 75\%.  We also note that sensitivity and specificity are evaluated separately in \cite{staaf08segmentation}, and this weakness has been perpetuated in all subsequent papers based  on the same evaluation framework.

\paragraph{A manually annotated data set \cite{hocking13learning}.}
The authors analyzed hundreds of neuroblastoma array-CGH profiles in order to define regions containing breakpoints (true signals), and regions not containing breakpoints (false signals).  This data set is freely distributed on CRAN\footnote{\url{http://cran.r-project.org/web/packages/neuroblastoma/}}.  Based on this large data set with known truth, the authors have performed a comprehensive comparison of segmentation methods for array-CGH data based on ROC curves.  A drawback of this evaluation framework is that once a particular data set is chosen, it is not possible to tune the signal-to-noise ratio of the problem.  Moreover, annotating a new data set is a challenging task, because it has to be large enough to contain a set of change-points that discriminate between competing segmentation methods.  

\paragraph{A simulation model \cite{mosen-ansorena12comparison}.}
The authors designed a complex simulation model to generate ``realistic'' copy-number profiles.  This model is implemented in the R package {\code{CnaGen}}, which is available from the authors' web page\footnote{\url{http://web.bioinformatics.cicbiogune.es/cnagen/}}.
The simulation model depends on 24 parameters{\footnote{\code{CnaGen} version 2.1.}}.  Some of them are directly driven by biological considerations,  such as the percentage of tumor cells in the sample or intra-tumor heterogeneity.  We empirically found it difficult to find a combination of parameters that yield realistic copy-number profiles. This may be due to the fact that the underlying data generation model is Gaussian.
Table \ref{tab:comp-data-generation} summarizes the features of approaches reviewed above.
\begin{table}[!h]
  \small
  \centering
  \begin{tabular}{r|c|c|c|c|c}
  Reference & \cite{willenbrock05a-comparison} &\cite{staaf08segmentation} & \cite{hocking13learning} & \cite{mosen-ansorena12comparison} & This paper \\ \hline
  Based on real biological data? & $\surd$ & $\surd$ & $\surd$& - & $\surd$\\
  Noise level based on a biological parameter? & - & $\surd$ & - & $\surd$ & $\surd$ \\
  Data generation possible? & $\surd$ & - & - & $\surd$ & $\surd$ \\
  Available as an R package? & $\surd$ & -  & $\surd$ & $\surd$ & $\surd$ \\
\end{tabular}
\caption{Features  of existing frameworks for real copy
  number data with known ``truth''. }
\label{tab:comp-data-generation}
\end{table}

\subsection{Proposed data generation mechanism}
\label{sec:prop-data-gener}
Based on these considerations, we propose an original data generation framework which aims at combining the advantages of all of the above-mentioned existing approaches.  Two necessary and sufficient ingredients for generating  a copy-number profile of length $n$ are:
\begin{itemize}
\item truth, in the form of $K$ breakpoint positions (out of $n-1$ intervals between two successive loci) and $K+1$ copy-number state labels for all $K+1$ regions between two consecutive breakpoints;
\item signal, in the form of locus-level data. For SNP arrays, this is generally a $n \times 3$ matrix of total copy numbers ($c$), allelic ratios ($b$), and germline genotypes.
\end{itemize}
Our proposed approach is described below. 

\subsubsection{Generation of ``truth''}
\label{sec:generation-truth}
\begin{figure}
\centering
\includegraphics[scale=0.60]{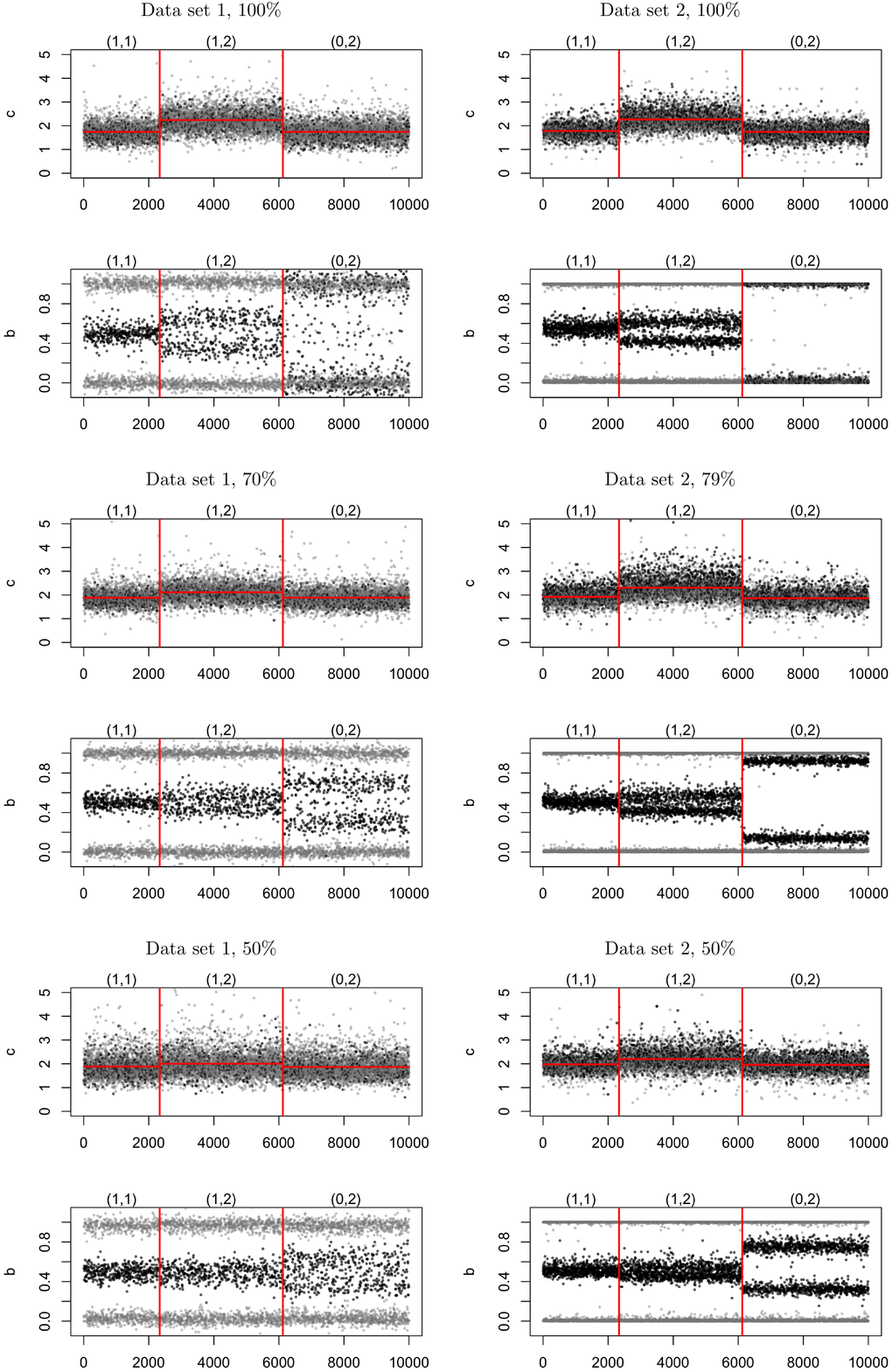}
\caption{Illustration of the variety of copy-number profiles that can be generated from the same ``truth'' as in Figure \ref{fig:copy-number-data-c-b}.  Each block of two plots corresponds to total copy numbers ($c$) and allelic ratios ($b$) for one particular combination of fraction of tumor cells (in rows) and data set  (in columns). Red vertical lines represent change points. SNPs that are \suppress{homozygous} \add{heterozygous} in the germline are colored in black; all of of the other loci are colored in gray.}
\label{fig:data-generation}
\end{figure}

When breakpoints and region labels are not user-supplied, we propose the following approach for generating them:
\begin{description}
\item[breakpoints:] given a signal length $n$, draw $K$ breakpoint positions uniformly out of the $n-1$ possible intervals between successive data points (vertical red lines in Figure~\ref{fig:data-generation});
\item[region labels:] draw $K+1$ region labels from a pre-defined set of copy-number state labels, such as normal, gain of one copy, hemizygous deletion, homozygous deletion, copy-neutral LOH (labels on top of each plot in Figure~\ref{fig:data-generation}).  By default, all region labels are equiprobable, but the user may provide a vector of probabilities for each desired region label.  
By default, successive regions are constrained in such a way that only one of the two parental copy numbers changes at the breakpoint.  
\add{Not adding such a constraint would be equivalent to allowing two distinct biological events to occur at the same genomic position, which is possible in theory but rarely observed in practice.} 
\end{description}


\subsubsection{Generation of locus-level data}
\label{sec:locus-level-data}
Given breakpoint positions and region labels, we generate a copy-number profile as follows: for each region of size $n_R$ between two breakpoints, we sample $n_R$ data points from a real copy-number data corresponding to this type of region.

The data generation mechanism therefore relies on real data where the underlying region label is (assumed to be) known.  We have made available two such ``real data sets with known truth'' in the package: each of them corresponds to a different SNP array platform (Affymetrix or Illumina), and both of them are taken from dilution series, consisting of mixtures of DNA from a tumor cell line and from blood cells originating from the same patient, with varying mixture proportions. For both data sets, we have selected several genomic regions which are representative of the diversity of copy-number states that are typically observed in tumor samples. Contrary to \cite{willenbrock05a-comparison}, these labels do not rely on any automatic segmentation or calling method.  Both data sets are described in Appendix~\ref{sec:snp-array-data}.

\subsection{Features of the proposed data generation mechanism}
\label{sec:feat-prop-data}
Our proposed data generation mechanism enjoys the following features:
\begin{itemize}
\item simplicity: small number of required parameters, all of which have a clear biological interpretation.  In particular, for a given data set, the noise level is governed by the fraction of tumor cells.  This is illustrated by Figure~\ref{fig:data-generation};
\item flexibility: the user may specify breakpoint positions and region labels directly, if desired.  Therefore, it is also possible to generate profiles with the same underlying ``truth'', but with different SNR, as illustrated by Figure~\ref{fig:data-generation};
\item reliability: copy-number regions were identified using the profiles with 100\% tumor cells.  In these profiles, the region labels may be defined manually unambiguously.  Because the same tumor cell line is used for the dilutions series from a given platform, the regions identified on the profiles with 100\% tumor cells can also be considered as ground truth for the profiles with less tumor cells, where direct manual identification would have been more problematic;
\item versatility: the design choice of separating ``truth'' generation from locus-level data generation implies that it is relatively easy to:
  \begin{itemize}
  \item annotate a new data set. Although dilution series are not publicly available for all possible platforms, it is also possible to annotate representative profiles from a given data set.  Moreover, annotating a new data set is not time-expensive, as one only needs to identify a few copy-number regions.
  \item extend the framework to other data types (for example array-CGH or high-throughput exome capture or whole genome sequencing) is straightforward: only a set of annotated data is required.
\end{itemize}
\end{itemize}

\section{Evaluation pipeline}
\label{sec:segm-pipel}
Now that we have a framework to generate data, we describe how to evaluate the performance of segmentation methods. 
\subsection{Benchmark}
\label{sec:segm-pipel-gen-dat}
Synthetic copy-number profiles were generated as described in Section~\ref{sec:generating-data}:
  \begin{description}
  \item[region-level ``truth'']: Each profile contains $n=200,000$ loci in copy number signal and $K=20$ breakpoints.  We chose to impose the constraint that on average, $90\%$ of segments are either normal (1,1), copy-neutral LOH (0,2), single copy-gain (1,2) or hemizygous deletion (0,1).  The remaining $10\%$ of regions are given less common copy-number states, such as homozygous deletion, or balanced duplication. \add{These parameters were inspired by our experience with SNP array data from The Cancer Genome Altas (TCGA), especially on ovarian cancers, where normal regions and regions of copy-neutral LOH, single copy-gain, and hemizygous deletion are fairly common, while other types of alterations are much more rare ~\cite{tcga11integrated}.}

  \item[locus-level data:] for each of $B=50$ such ``truth'' profiles, corresponding locus-level data are then generated for 100\%, 70\% and 50\% of tumor cells for data set 1, and 100\%. 79\% and 50\% of tumor cells for data set 2. \add{These percentages are among those available from the dilution series from which real data was extracted, see Appendix~\ref{sec:snp-array-data}.  Pure tumor samples (100\%) are typically observed in studies about tumor cell lines, while percentages as low as 50\% are typically observed in primary tumors.}
  \end{description}

\subsection{Preprocessing}
We $\log$-transformed total copy numbers to stabilize their variance and smoothed outliers using {\code{smooth.CNA}} \cite{olshen04circular} as it improved segmentation results for all methods. Allelic ratios were converted to (unimodal) decrease in heterozygosity ($d$) as described in Section \ref{sec:dna-copy-number}.

\subsection{Compared segmentation methods}
\label{sec:comp-segm-meth}
We evaluated different types of methods belonging to the different classes described in Section~\ref{sec:typology-copy-number}: multiple change-point, recursive, fused, and HMM-based methods. These methods are described in Table \ref{tab:method-summary}, where we mention which of them are able to process both signal dimensions ($c$ and $d$) or only one of them.  
\begin{table}[!h]
  \centering
  \begin{tabular}{p{1cm}p{1.8cm}p{3.5cm}lccl}
  \hline
  &&&& \multicolumn{2}{c}{Time (s)}  & \\
Name & {\code{R package}} & {\code{function}}& dims & n=$10^4$ & n=$10^5$  & Ref \\ 
  \hline
  \multicolumn{7}{c}{Multiple change-point}\\
DP & {\code{cghseg}} & {\code{segmeanCO}} & 1d & 0.24 & 2.37 & \cite{rigaill10pruned} \\
CST & {\code{cnaStruct}} &{\joint{\code{segment}}}& 2d& 120 & fail & \cite{MosenBivariate2013} \\
DP & {\code{jointseg}} & \joint{\code{doDynamicProgramming}}  & 2d & 140 & fail  \\ 
\\
  \hline
  \multicolumn{7}{c}{Recursive}\\
CBS  &  {\code{DNAcopy}} & {\code{segment}} & 1d & 0.34 & 1.69 & \cite{venkatraman07a-faster} \\
PSCBS & {\code{PSCBS}}&{\joint{\code{segmentByPairedPSCBS}}} & 2d &  1.04 & 4.00  & \cite{olshen11parent-specific}\\
RBS & {\code{jointseg}}& {\joint{\code{doRBS}}}  & 2d & 0.15 & 1.15  & \cite{gey08using}\\
\\
  \hline
  \multicolumn{7}{c}{Fused}\\
GFLars & {\code{jointseg}}&{\code{doGFLars}}  & 1d & 0.29 & 3.70   & \cite{harchaoui08catching}\\
GFLars & {\code{jointseg}}&{\joint{\code{doGFLars}}} & 2d  & 0.08 & 0.60 & \cite{bleakley11the-group}\\
\\
 \hline
 \multicolumn{7}{c}{HMM}\\
PSCN & {\code{PSCN}}& {\joint{\code{segmentation}}} & 2d & 7.25 & 73  & \cite{chen11estimation}\\
  \hline
  \end{tabular}
  \caption{List of DNA copy number segmentation methods evaluated.}
  \label{tab:method-summary}
\end{table}
Not all of these methods were implemented in {\code{R}}.  We ported  from {\code{Matlab}} {\code{GFLseg}}\footnote{Available at \url{http://cbio.ensmp.fr/~jvert/svn/GFLseg/html}.} to {\code{R}} the implementation of  multi-dimensional dynamic programming and the group-fused LARS \cite{bleakley11the-group}, and we implemented recursive binary segmentation \cite{gey08using} in {\code{R}}.  In practice, as recommended by \cite{gey08using,harchaoui08catching,bleakley11the-group}, both group-fused LARS and recursive binary segmentation are used to quickly identify a list of \emph{candidate} change points, which is then pruned using dynamic programming.

\add{All of the compared methods are reasonably fast and memory-efficient, except those based on two-dimensional dynamic programming (DP): {\code{cnaStruct}} and our implementation of DP in {\code{R}} . Indeed,  two-dimensional DP is quadratic in time and memory and thus cannot handle profiles of size $n=10^5$.  It may be surprising that the two-dimensional version of GFLars is faster than its one-dimensional counterpart.  This is a consequence of the fact that the number of informative markers is several times larger for $(c)$ than for $(d)$ (as explained in Section \ref{sec:dna-copy-number}).  As the implementation of GFLars does not handle missing values, the 2d version of GFLars was applied to non-missing entries in $(c,d)$, while the 1d version was applied to a much longer signal (all $(c)$ entries).  This phenomenon does not happen for other two-dimensional segmentation methods as their implementation does handle missing values.}

\subsection{Criteria for performance evaluation}
\label{sec:crit-perf-eval}
Comparison studies typically assess the performance of DNA copy number analysis methods either in terms of their ability to accurately identify breakpoint locations~\cite{lai05comparative,hocking13learning}, copy-number states~\cite{staaf08segmentation,mosen-ansorena12comparison}, or both~\cite{willenbrock05a-comparison}.  This paper focuses on the former only, because we are interested in comparing segmentation methods.  The problem of evaluating strategies for calling copy-number states is left for future work.

\begin{figure}
 \centering
 \includegraphics[width=\columnwidth]{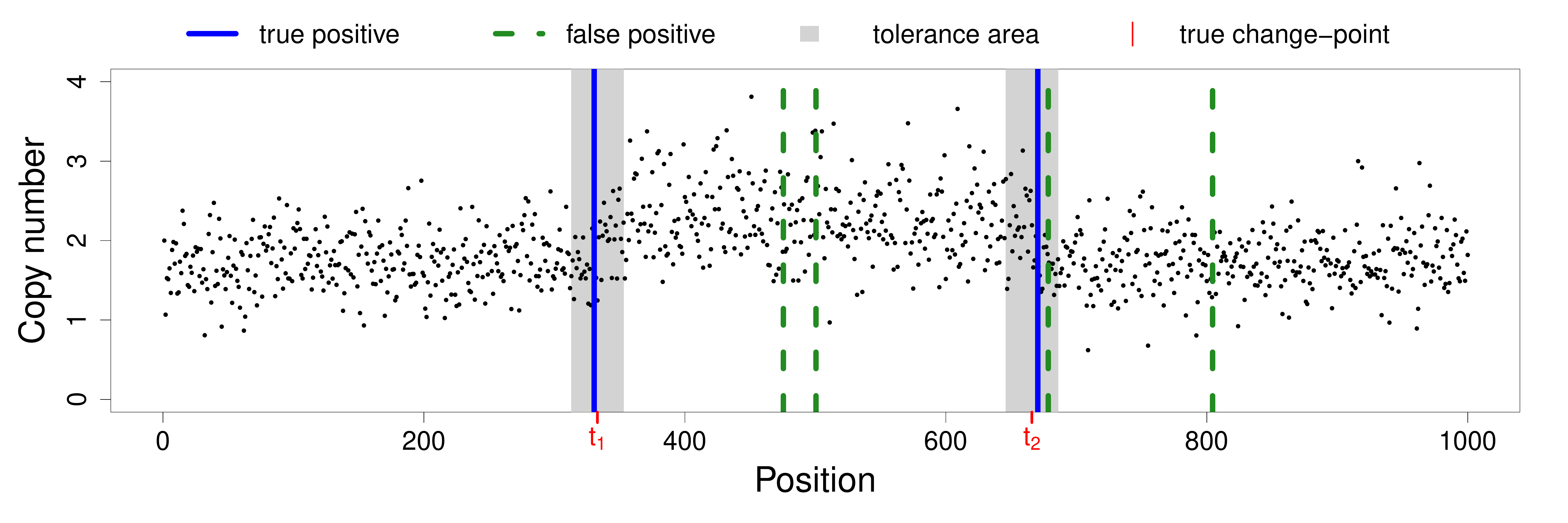}
 \caption{Definition of false positive and true positive to build performance evaluation.}
 \label{fig:TPTN}
\end{figure}

As our proposed data generation framework provides copy number profiles with known ``truth'', a natural way to evaluate the performance of a given method is to cast the problem of breakpoint detection as a binary classification problem.  Specifically, for each generated copy number profile, we know where the true breakpoints are located.   The number of true positives TP  is the number of true breakpoints for which at least one breakpoint is detected closer than a given tolerance parameter.  The number of false positives FP is defined as FP=P-TP, where P is the number of ``positives'', that is, the total number of detected breakpoints.  With this definition, whenever a method identifies two or more breakpoints within the tolerance area of a true breakpoint, one of these breakpoints counts as a true positive, while all others count as false positives.   
This definition of true and false positives is illustrated by Figure \ref{fig:TPTN}, where gray areas highlight tolerance areas around the true change-points, whose positions are identified as $t_1$ and $t_2$ on the $x$ axis. In this example, breakpoints were detected in both shaded areas, therefore the number of true positives (solid blue lines) is two.  There are four false positives (dashed green lines):  one in a gray area where there is already one true positive, and three which are not within the tolerance area of any true breakpoint.
 \add{Alternative definitions of true and false positives may be considered.  Some of these alternatives are implemented in the {\code{jointseg}} package, including one in which a second breakpoint found within a tolerance area is not counted as a false positive.  We chose to stick with the above-described evaluation (where such breakpoints are called false positives) in order not to favor methods such as the (group) fused lasso that tend to systematically find multiple breakpoints very close to each other, which is generally inconsistent with the biology of cancers.}

\paragraph{Related works.} A similar definition of true and false positives is used in \cite{willenbrock05a-comparison}, although the authors do not mention how the above case of multiple breakpoints within the tolerance area is handled.  Another related approach has been proposed in \cite{hocking13learning}.  There, copy-number profiles are real, array-CGH profiles for which regions containing a breakpoint and regions containing no breakpoints have been delineated by experts.  The main difference is that only a subset of the ``true'' and ``false'' breakpoints are annotated, and that the tolerance parameter cannot be tuned without the expert re-annotating the data set.  Finally, a similar type of evaluation has been used by \cite{lai05comparative}, at the the locus level instead of the breakpoint level. This locus-level based evaluation method tends to favor segmentation methods that accurately identify large altered regions, even if they fail to detect breakpoints delineating smaller altered regions. 

\subsection{ROC-based evaluation}
Usually, each method provides a segmentation and its associated set of breakpoints. This can be translated into a measure of sensitivity and specificity using the above definition of true and false positives. However, the methods have to be compared at the same specificity or sensitivity level in order for this comparison to be fair. 
Ideally, we would like to compute a Receiver Operator Characteristic (ROC) curve for each method.  In order to do this, one needs to explore a large set of possible segmentations with varying sensitivity and specificity, obtained by exploring the set of tuning parameters of each method.  Such an exhaustive exploration is tedious and time consuming as soon as the number of parameters is larger than 2 or 3, and may lead to over-optimistic results.  To overcome this problem, we adopted the following strategy: for any given method $m$, we recovered a segmentation in $k_m$ change points using default parameters, and we retrieved for each $k \in \{1 \dots k_m\}$ the best $k$ subset of these $k_m$ using dynamic programming.  Another possible strategy would be to sort the $k_m$ change points according to a measure of confidence. 

One could be worried that the range of explored sensitivity/specificity is highly variable across methods. In practice, our experience is that the default parameters of a method generally tend to over-segment the data and that typically, most of the true change points are found, at the cost of a more or less large number of false positives. This is in agreement with \cite{hocking13learning}. 

\section{Results}
\label{sec:results}

\subsection{Quantifying problem difficulty for known change points}
\label{sec:problem-difficulty}
Segmentation methods rely on a statistic to quantify the biological difference between any two regions.  Based on this statistic, they aim at locating a good set of regions or equivalently, of change points. This location problem is combinatorial in nature.  In this section, we try to quantify this biological difference independently of this combinatorial problem.  
In order to do this, we assume that change point positions are given \textit{a priori} and we compare the power to call a change using total copy numbers ($c$) or allelic signals  ($d$) for different types of change points. In order to perform this power study, we need to formally define the notion of power, or signal-to-noise ratio (SNR), between copy number regions.  We chose a definition of SNR which is consistent with our proposed data-generation mechanism, in which DNA copy number data from a given region are sampled from a population which represents the corresponding copy-number state (see Section~\ref{sec:feat-prop-data}).  Let us consider two regions and label by ``0'' and ``1''  the copy number state of two regions.  For univariate signals ($c$ or $d$), a natural definition of SNR is the (squared) $Z$ statistic of the comparison between the sample means of region ``0'' and region ``1'':
\begin{eqnarray}
\label{eq:snr-c}
\snr(c) = \frac{\left(\bar{c}_0-\bar{c}_1\right)^2}{\sigma_{c,0}^2/n_0+\sigma_{c,1}^2/n_1}\\
\label{eq:snr-d}
\snr(d) = \frac{\left(\bar{d}_0-\bar{d}_1\right)^2}{\sigma_{d,0}^2/n^{\star}_0+\sigma_{d,1}^2/n^{\star}_1} \,,
\end{eqnarray}
where $n_i$ is the total number of loci in region $i$, $\bar{c}_i$ and  $\sigma_{c,i}$ are the sample mean and population standard deviation of total copy numbers in state $i$ and $\bar{d}_i,\sigma_{d,i}$ are the sample mean and population standard deviation of the decrease in heterozygosity in state $i$.  Note that the decrease in heterozygosity is only defined for SNPs that are heterozygous in the germline, whereas the total copy number is defined for all loci. Therefore, $\bar{d}_i$ is calculated based on $n^{\star}_i$ heterozygous SNPs, while $\bar{c}_i$ is calculated based on all $n_i$ loci.  For a given DNA sample, the fraction of heterozygous SNPs among those present on the microarray is typically close to 1/3; moreover, data set 1 contains not only SNP probes but also non-polymorphic loci, with a 1:1 ratio.  As a result, the fraction $n^{\star}_i/n_i$ is approximately $1/6$ for data set 1 and $1/3$ for data set 2.
A natural extension of this definition of SNR to the two-dimensional case of the statistic $(c,d)$ is 
\begin{eqnarray}
  \label{eq:SNR-c-d}
  \snr(c,d) = \left(\bar{c}_0-\bar{c}_1, \bar{d}_0-\bar{d}_1\right) \left(S_0+ S_1 \right)^{-1} \left(\bar{c}_0-\bar{c}_1, \bar{d}_0-\bar{d}_1\right)' \,,
\end{eqnarray}
where $S_i$ is the population covariance matrix of the bivariate vector $(c,d)$, that is $S_i=\begin{pmatrix}
   \sigma_{c,i}^2/n_i & \tau_{cd,i}/n^{\star}_i \\
   \tau_{cd,i}/n^{\star}_ i& \sigma_{d,i}^2/n^{\star}_i
\end{pmatrix}$ with $\tau_{cd,i}$ the covariance between $c$ and $d$ in state $i$.
In practice, the population parameters for copy-number state $i$ (that is, $\sigma_{d,i}$, $\tau_{cd,i}$,  and $\sigma_{d,i}$) are calculated from the annotated data. The sample parameters ($\bar{c}_i$ and $\bar{d}_i$) are calculated from samples of $n_i$ and $n^{\star}_i$ loci, respectively.  Note that $\snr(c)$ and $\snr(d)$ are comparable with each other since they follow (non-centered) $\chi^2$ distributions with 1 degree of freedom under the null hypothesis of no breakpoint between state $0$ and state $1$.

By definition, SNR is an increasing function of the length of each flanking segment.  For $i \in \{0,1\}$, we chose $n_i=500$.  $n^{\star}_i$ depends on \add{the} proportion of heterozygous SNPs in the sample; as explained above, it is very close to $n_0/6$ for data set 1 and  $n_0/3$ for data set 2. Therefore, the length of the flanking regions essentially acts as a constant scaling factor across all transitions and settings.  Therefore, SNR only reflects differences between the underlying copy number states. 
Figure~\ref{fig:SNR-data-pop} shows the average (and standard error) of $\log(\snr)$ across 100 samplings for three levels of tumor purity level, for three common types of copy number transitions  for data set 1 (top panel) and data set 2 (bottom panels).  Several conclusions may be drawn: 
\begin{figure}
  \centering
  \begin{tabular}{rcccl}
  & normal/gain & normal/deletion & loss/copy-neutral LOH\\
  \raisebox{3.5em}{
    \rotatebox{90}{$\log(\snr)$}
  } 
  & \includegraphics[width=.27\textwidth,trim=30 10 30 60, clip]{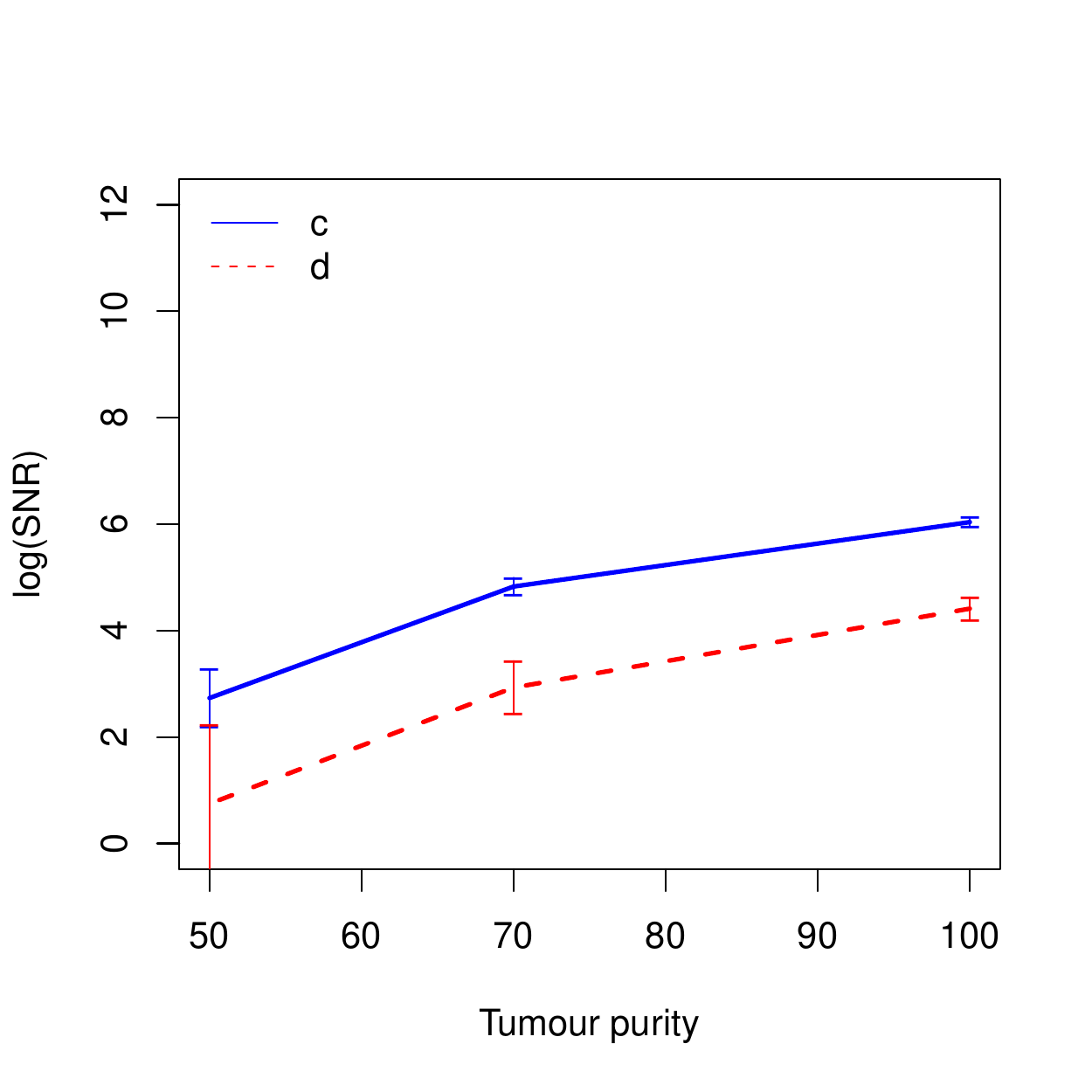}
  & \includegraphics[width=.27\textwidth,trim=30 10 30 60, clip]{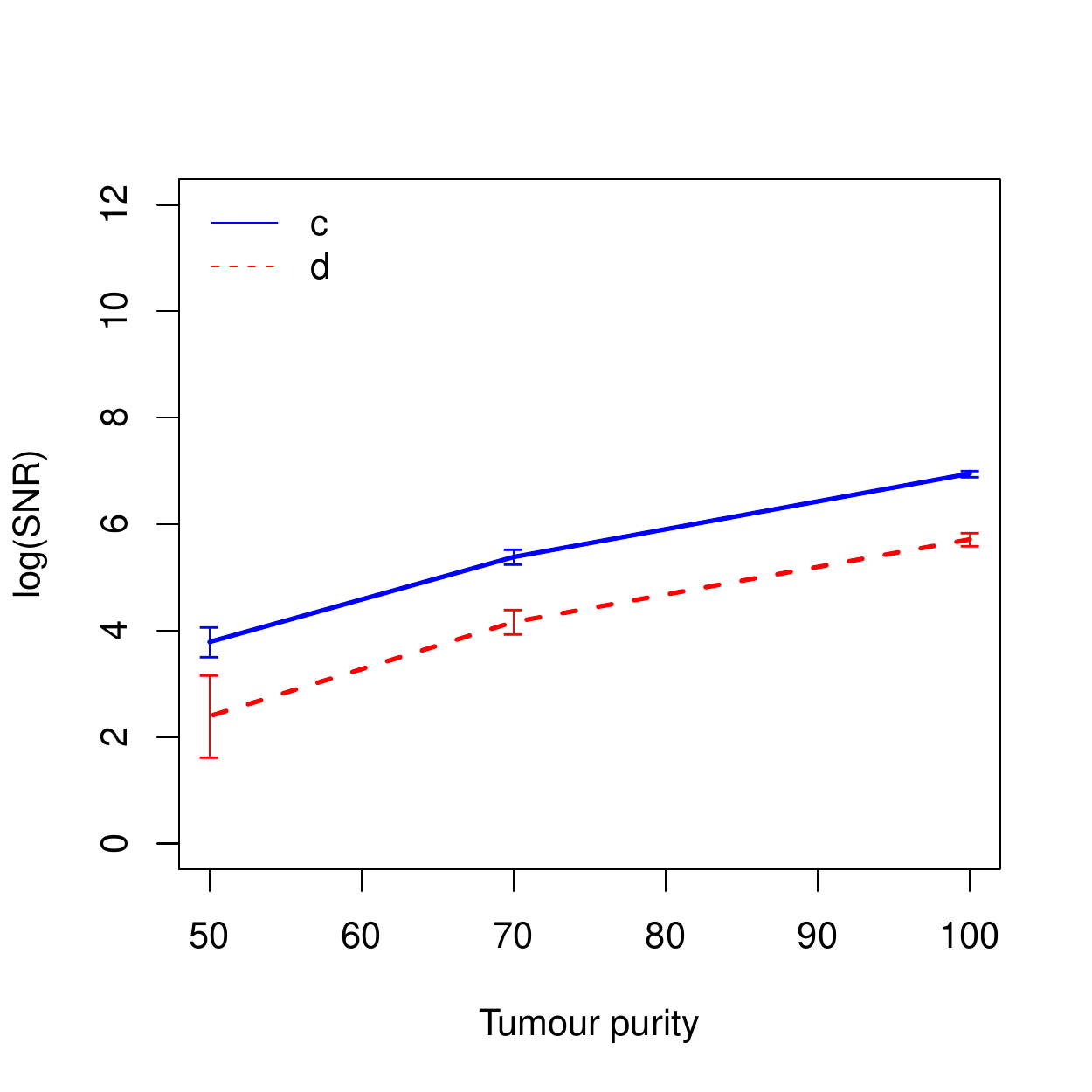}
  & \includegraphics[width=.27\textwidth,trim=30 10 30 60, clip]{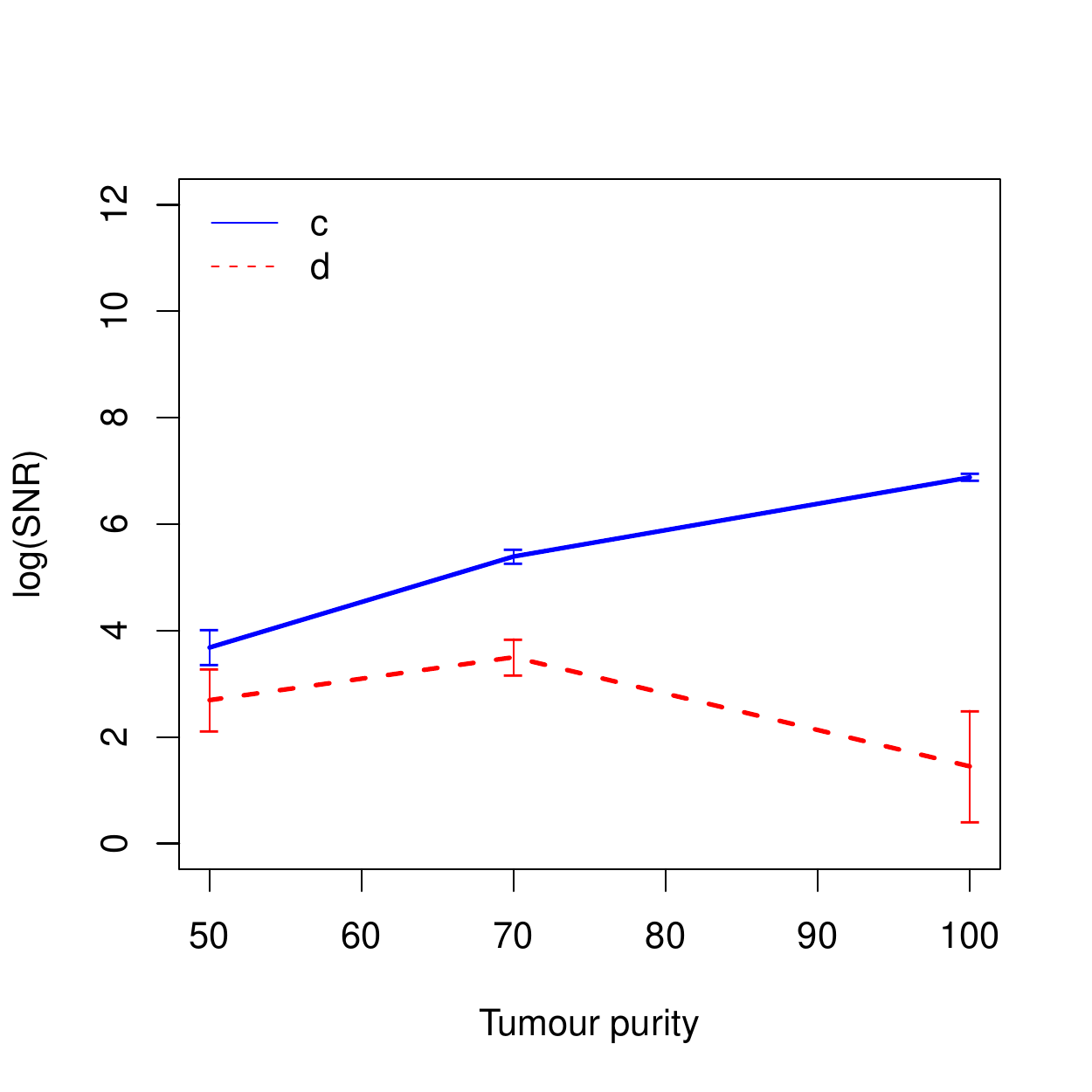} 
  & \raisebox{3em}{
  \rotatebox{90}{Data set 1}
  } \\
  
  \raisebox{3.5em}{
    \rotatebox{90}{$\log(\snr)$}
  } 
  & \includegraphics[width=.27\textwidth,trim=30 10 30 60, clip]{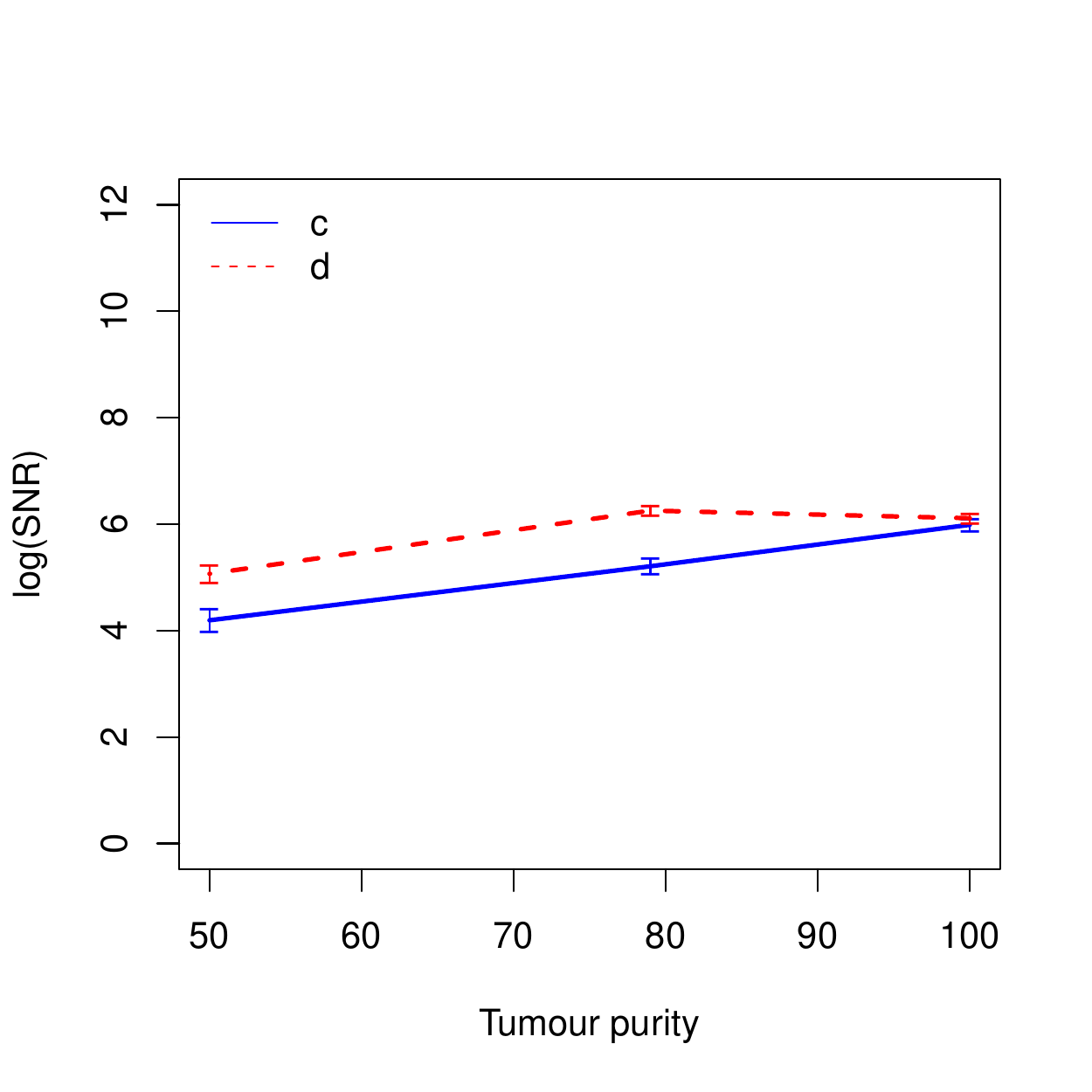}
  & \includegraphics[width=.27\textwidth,trim=30 10 30 60, clip]{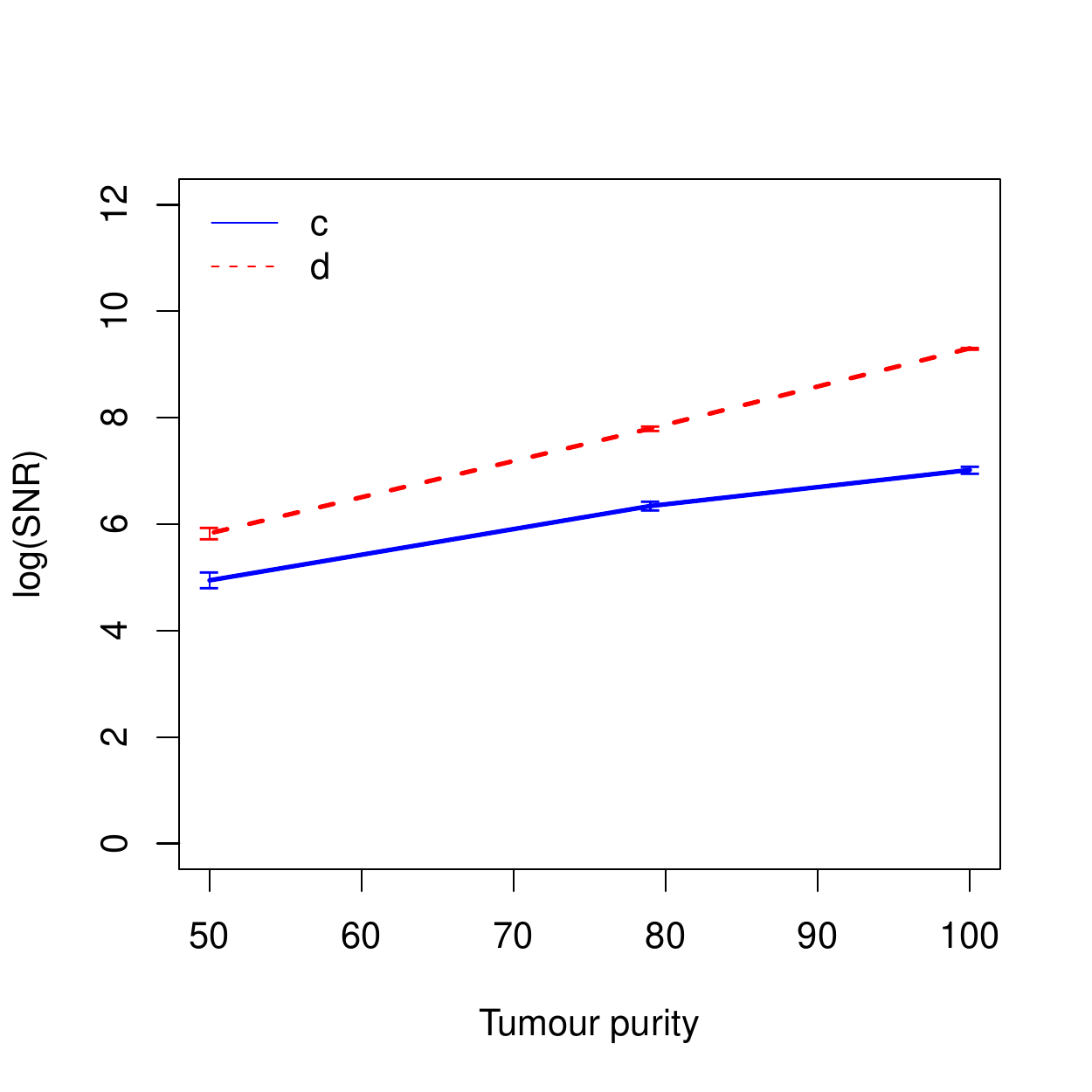}
  & \includegraphics[width=.27\textwidth,trim=30 10 30 60, clip]{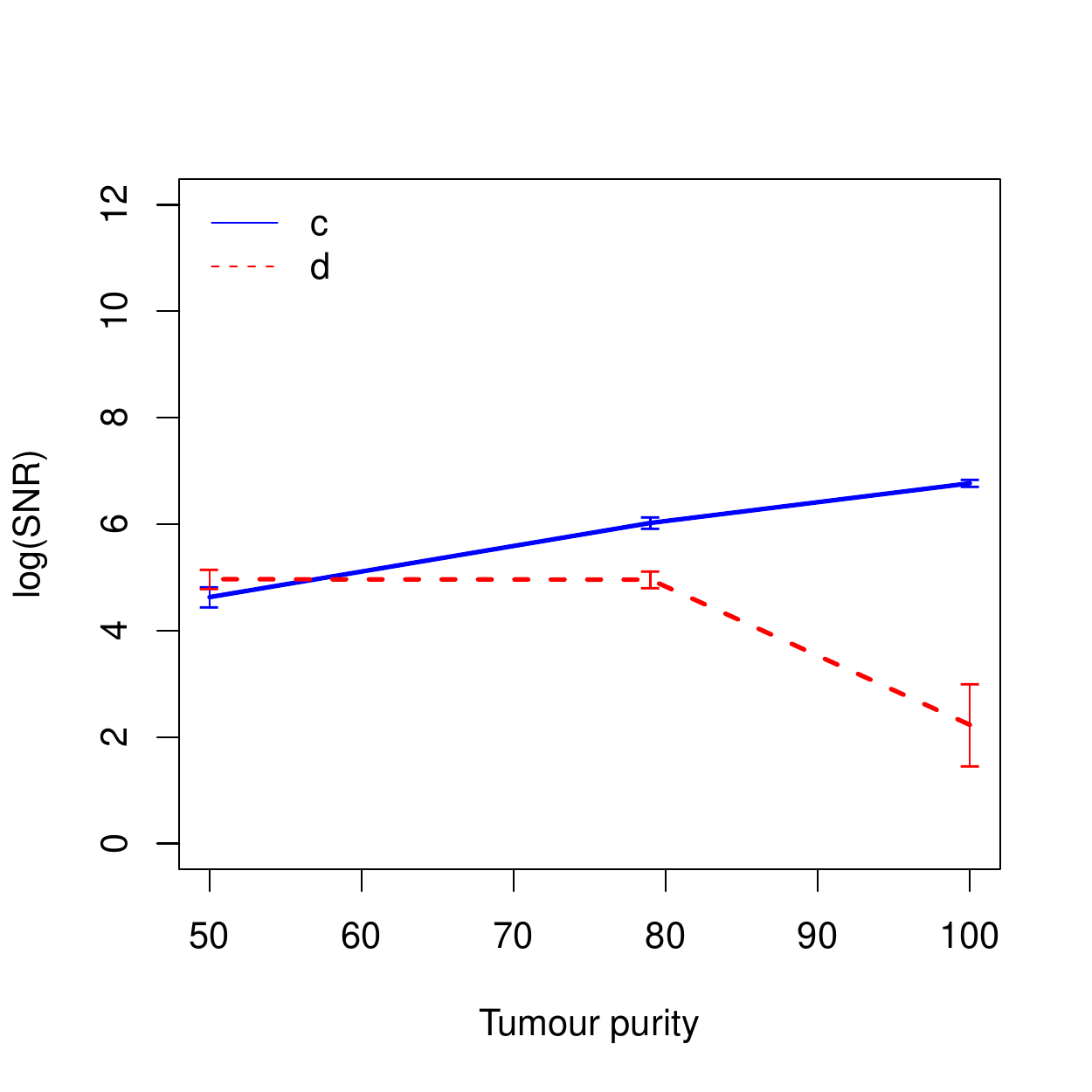}
  &  \raisebox{3em}{
    \rotatebox{90}{Data set 2}
  } 
\end{tabular}
  \caption{Average $\log(\snr)$ and corresponding standard errors across 100 samples as a function of the percentage of tumor cells for total copy numbers ($c$, solid blue lines) and allelic ratios ($d$, dashed red lines). Each column corresponds to a type of copy number transition. Each row corresponds to a given data set.}
\label{fig:SNR-data-pop}
  \end{figure}

\begin{itemize}
\item \textbf{Difficulty generally increases with normal contamination}: SNR generally increases with the percentage of tumor cells.  This is true for all types of transitions for $c$. For $d$, the only situation in which SNR is not an increasing function of tumor purity is the case of transitions between loss and copy-neutral LOH (Figure  \ref{fig:SNR-data-pop}, rightmost column).  This is expected theoretically because both of these states correspond to LOH in the tumor cells of the sample, implying that the true $d$ in these cells is 1.  In presence of normal cells, $d$ estimates in both states are shrunk $d$ toward 0, but in a state-specific way (see \cite[Figure 4]{bengtsson10tumorboost} for a detailed explanation of this phenomenon);
\item \textbf{SNR levels depend on the type of copy number transition} for a given data set (that is, for a given row in Figure~\ref{fig:SNR-data-pop}).  This holds for both statistics ($c$ or $d$).  Note that in the case of $c$, this is unexpected, as all plotted transitions correspond to a one-copy gain.
\item \textbf{Possibly low power.} Note that in some cases (e.g. data set 1, (a) and (c)), the computed SNR is lower than 2.  Under the null hypothesis of no difference in mean levels, SNR follows a centered $\chi^2(1)$ distribution, so that this range of observed SNR correspond to $p$-values of the order of 1\%, which is not low considering the large number of data points ($n_i=500$).
\item \textbf{Neither $c$ or $d$ is always the best statistic}. For a given type of transition (that is, for a given column in Figure~\ref{fig:SNR-data-pop}) and a given statistic, the trend in SNR is comparable across data sets.  However, the relative power of $c$ with respect to $d$ is much higher for data set 1 than for data set 2.  This is directly related to the above-mentioned difference between ratios $n^\star_i/n_i$ of the number of informative loci for each statistic.
\end{itemize}

In this subsection, we assessed the intrinsic difficulty of calling a change point if the positions to test are known \textit{a priori}. This study suggests that $c$ and $d$ are complementary sources of information, implying that change point detection methods should ideally take both of them into account.  This study also sheds light on the fact that low percentages of tumor cells severely impacts SNR.  In the remaining subsections, we assess the ability of segmentation methods to recover the true location of change points. 

\subsection{Robustness of the evaluation to the tolerance parameter}
\label{sec:tol-par-in}
Our first goal was to check the influence of the tolerance parameter on the methods' performance. Our simulations were run using data generation as described in section \ref{sec:segm-pipel-gen-dat}. We computed partial areas under the ROC curves (pAUC) with a number of false positives between 0 and 10.  Mean and 95\% confidence intervals of pAUCs across simulation runs were calculated for each method for 5 values of the tolerance parameter (1, 2, 5, 10 and 20).  For example, a tolerance of 5 means that a breakpoint is considered correct if it lies within 5 data points of the true breakpoints (see section~\ref{sec:crit-perf-eval} for more details). These results are reported in Figure \ref{fig:Affy-AUC,tolInfluence} in the scenario without normal contamination.  Similar results were observed for other scenarios. 

\begin{figure}
  \centering 
  \includegraphics[width=.20\columnwidth]{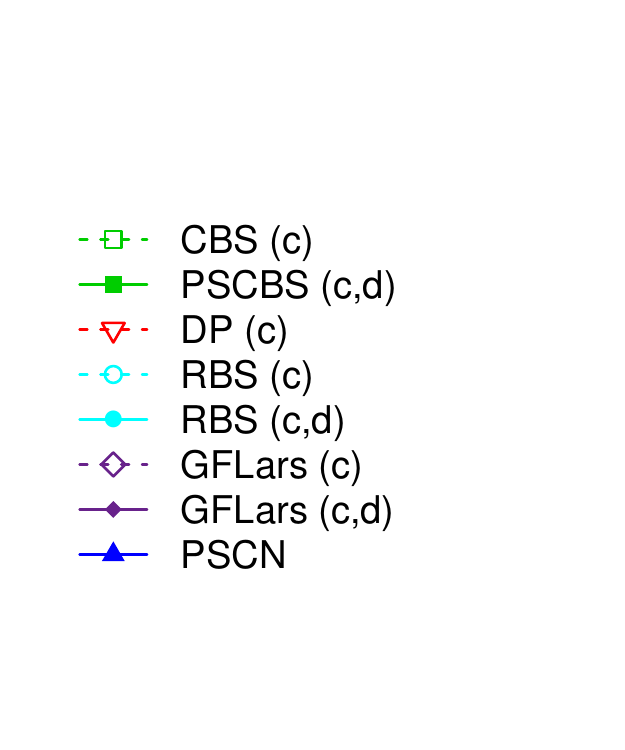}
  \includegraphics[width=.38\columnwidth]{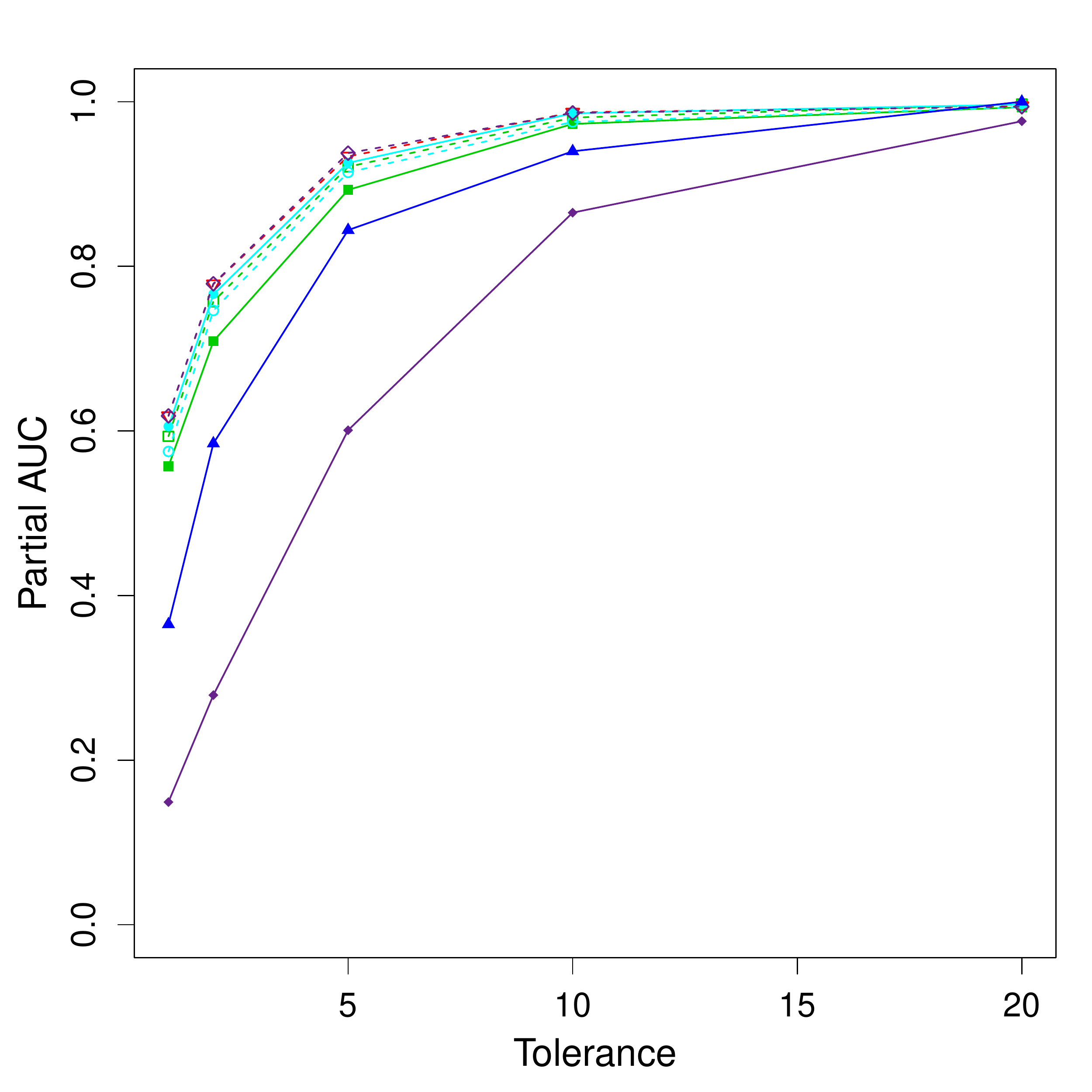}
  \includegraphics[width=.38\columnwidth]{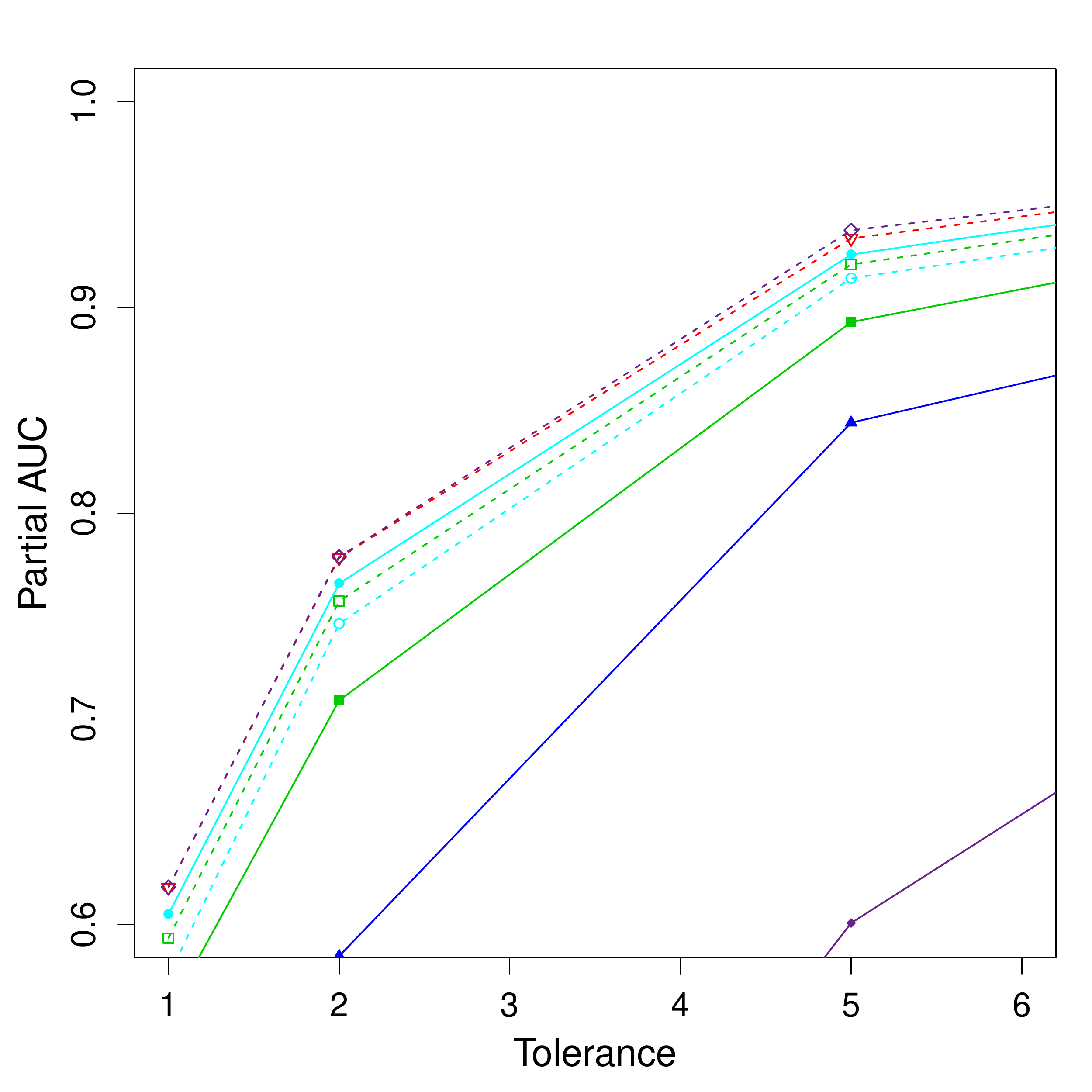}
  \caption{Method performance increase with the tolerance parameter for both data sets. Partial AUC for FP $\leq 10$ for data set 1 and 100\% tumor cells.}
  \label{fig:Affy-AUC,tolInfluence}
\end{figure}

Increasing tolerance clearly increases pAUC for all methods. This is the case even in the arguably ``simple'' scenario where no normal cells are present.  However, in most cases the ranking of all methods is not affected by tolerance. Based on these results, we decided to report only pAUC for one particular value of tolerance: 5 loci on each side of the breakpoints.

\subsection{Joint segmentation generally increases performance}
\label{sec:taking-both-dimens}
This section aims at comparing the quality of segmentations  obtained using total copy numbers only $(c)$, allelic ratios only $(d)$, and both of them $(c,d)$ and how the quality of the segmentation is affected by the purity of the sample. As explained in section~\ref{sec:problem-difficulty}, it is typically expected that localization of the breakpoints is easier using both dimensions of the signal.
In order to do so, we compared 6 scenarios corresponding to two data sets and three levels of purity (high, intermediate and low).
Table \ref{table:best-affy-illu} reports the pAUC of the best $(c)$, $(d)$ and $(c, d)$ methods for data set 1 and 2, respectively.
Detailed results for all methods are presented in Table \ref{table:all-affy-illu}.

\suppress{For both data sets it is quite clear that performance in terms of pAUC severely deteriorates when the level of contamination increases. Interestingly, $(c)$  methods perform better than $(d)$ methods for both high and intermediate level of purity.  For example in the case of data set 2 the minimum difference in pAUC between $(c)$ and $(d)$ is 26\% for high level of purity and 7\% for intermediate level of purity (Table \ref{table:all-affy-illu}). For a low level of purity, the results depend on the data set.
For data set 1, $(c)$ outperforms  $(d)$ but overall the pAUC are very low. For data set 2, $(d)$ outperforms $(c)$ with a minimum pAUC difference of 7\%. These observations are in agreement with the results of Section \ref{sec:problem-difficulty}. This difference between data sets 1 and 2 can be explained by the fact that the proportion of informative probes for $(d)$ (heterozygous SNPs) is twice less for data set 1 than for data set 2. 
Interestingly, $(c,d)$ methods do not always outperform $(c)$-only and $(d)$-only methods. However, as can be seen in Table \ref{table:best-affy-illu}, there are always several $(c,d)$ approaches among top performers. }

\add{For both data sets it is quite clear that performance in terms of pAUC severely deteriorates when the level of contamination increases. $(c)$  methods perform better than $(d)$ methods for high level of purity.  For example in the case of data set 2 the minimum difference in pAUC between $(c)$ and $(d)$ is 19\% for high level (Table \ref{table:all-affy-illu}). For an intermediate level of purity, for data set 1 $(c)$ outperforms $(d)$ with a minimum pAUC difference of 41\% and for data set 2 $(c)$ is similar to $(d)$. 
For a low level of purity, the pAUCs are low or very low for both data sets;
for data set 1, $(c)$ outperforms  $(d)$ with a minimum pAUC difference of 6\%; for data set 2, $(d)$ outperforms $(c)$ with a minimum pAUC difference of 15\%. These observations are in agreement with the results of Section \ref{sec:problem-difficulty}. The difference between data sets 1 and 2 can be explained by the fact that the proportion of informative markers is different, namely around 1/6 and 1/3, respectively. 
This low proportion of informative markers also explains the poor performance of GFLars $(c,d)$ (which could also be seen in Figure~\ref{fig:Affy-AUC,tolInfluence}), as the current implementations of 2d GFLars do not handle missing values in one of the dimensions.}

\add{Not all $(c,d)$ methods outperform $(c)$-only and $(d)$-only methods. For example, for data set 1 and 100\%, although PSCBS has good performance, it is outperformed by 2 to 5 \% by all $(c)$ methods. However, as can be seen in Table \ref{table:best-affy-illu}, there are always several $(c,d)$ approaches among top performers. }


\subsection{Choosing the appropriate method for a given context}
\label{sec:no-meth-uniform-best}
In practice, when analyzing SNP array data, biostatisticians and bioinformaticians will choose one particular method to perform data segmentation. This choice is often \textit{ad hoc} and based on personal experience. 
Our purpose here is not to make a comparison of all existing segmentation methods, but to compare relevant candidates in different classes of approaches. In the settings that we have considered it seems that RBS $(c,d)$ performs very well. However, the point of our framework is not to select once and for all a best segmentation tool, but rather to justify the use of one method for one particular type of scenario (cancer type, cellularity, data set). In particular, we make no claim about the performance of RBS for other data sets.

\begin{table}[ht]
\centering
\begin{tabular}{crrrp{0.1cm}rrr}
  \hline
 & \multicolumn{3}{c}{Data set 1} & & \multicolumn{3}{c}{Data set 2}\\
  \cline{2-4}\cline{6-8}
  Statistic &  100\% & 70\% & 50\% & & 100\% & 79\% & 50\% \\ 
  \hline
  $(c,d)$ & 0.93 & 0.63 & 0.22 & & 0.97 & 0.95 & 0.75 \\ 
  $(c)$ & 0.94 & 0.64 & 0.18 & & 0.96 & 0.89 & 0.49 \\ 
  $(d)$ & 0.35 & 0.18 & 0.10 & & 0.71 & 0.84 & 0.67 \\
   \hline
\end{tabular}
\caption{Best pAUC across methods for each combination of statistic, data set and percentage of tumor cells.}
\label{table:best-affy-illu}
\end{table}

\begin{table}[ht]
\centering
\begin{tabular}{clrrrp{0.1cm}rrr}
  \hline
  & &\multicolumn{3}{c}{Data set 1} & & \multicolumn{3}{c}{Data set 2}\\
  \cline{3-5}\cline{7-9}
  Statistic & Method& 100\% & 70\% & 50\% & & 100\% & 79\% & 50\% \\ 
  \hline 
  \multirow{3}{*}{$(c,d)$}&PSCBS & 0.89 & 0.60 & 0.16 & & 0.97 & 0.88 & 0.51 \\ 
  &GFLars & 0.60 & 0.42 & 0.14 & & 0.97 & 0.91 & 0.60 \\
  &RBS & 0.93 & 0.63 & 0.22 & & 0.97 & 0.95 & 0.75 \\ 
  \hline
  \multirow{3}{*}{$(c)$}&CBS& 0.92 & 0.59 & 0.16 & & 0.91 & 0.84 & 0.45 \\ 
  &GFLars & 0.94 & 0.64 & 0.18  &  & 0.96 & 0.89 & 0.49 \\ 
  &RBS & 0.91 & 0.62 & 0.17 & & 0.90 & 0.84 & 0.48 \\ 
  &cghseg& 0.93 & 0.61 & 0.18 & & 0.95 & 0.88 & 0.49 \\ 
  \hline
  \multirow{3}{*}{$(d)$}&CBS& 0.35 & 0.17 & 0.10 & & 0.71 & 0.83 & 0.64 \\ 
  &GFLars& 0.35  & 0.18 & 0.10 & & 0.71 & 0.84 & 0.66 \\ 
  &RBS & 0.34  & 0.17& 0.09 & & 0.69 & 0.83 &0.65  \\
  &cghseg& 0.35  &0.18 &0.10 & & 0.70& 0.84 & 0.67 \\ 
   \hline
\end{tabular}
\caption{pAUC by for each combination of method, statistic, data set and percentage of tumor cells.}
\label{table:all-affy-illu}
\end{table}

\subsection{Heterogeneity of breakpoint detection difficulty}

\begin{figure}
\centering
\includegraphics[width=0.49\columnwidth]{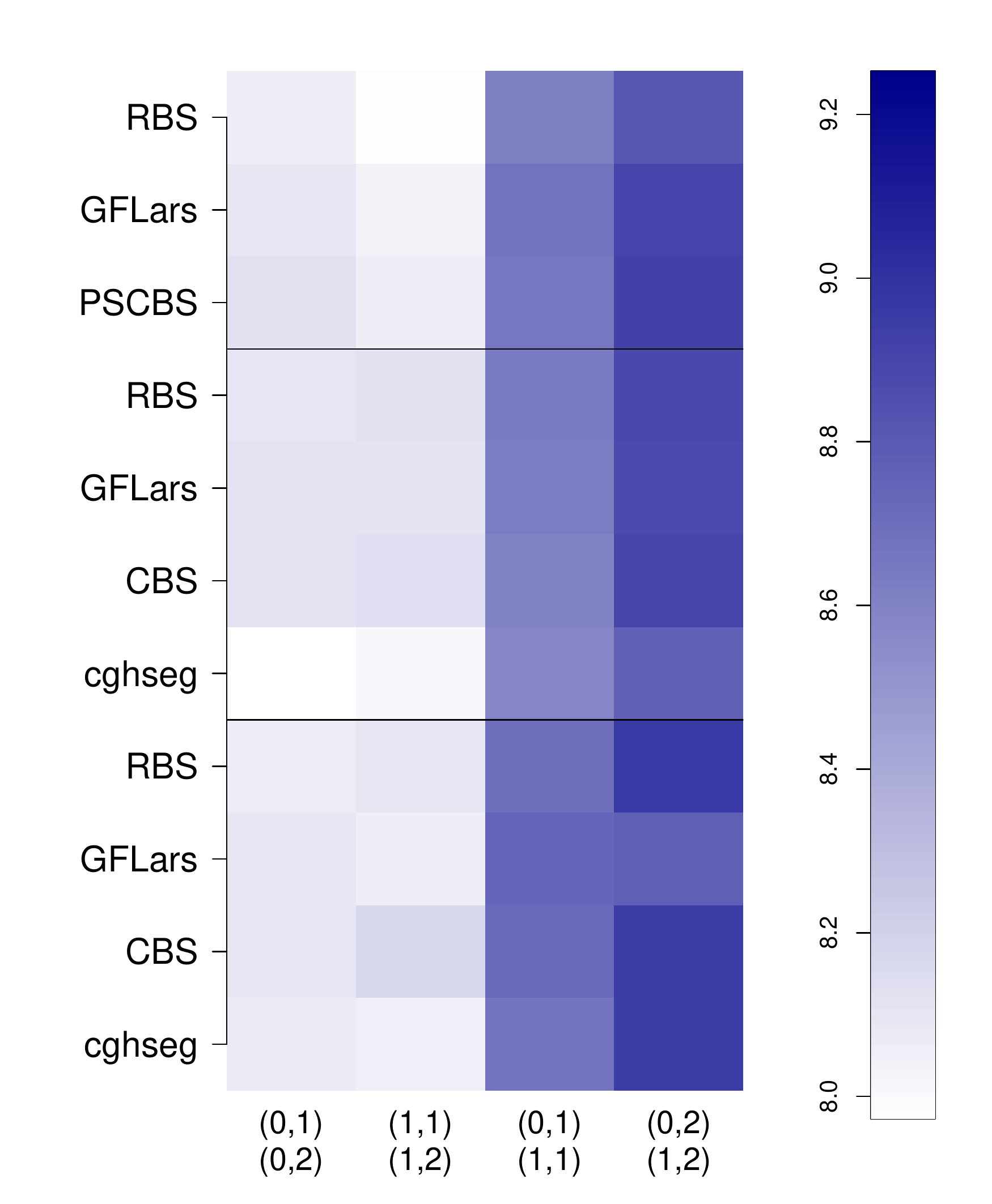}
\centering
\includegraphics[width=0.49\columnwidth]{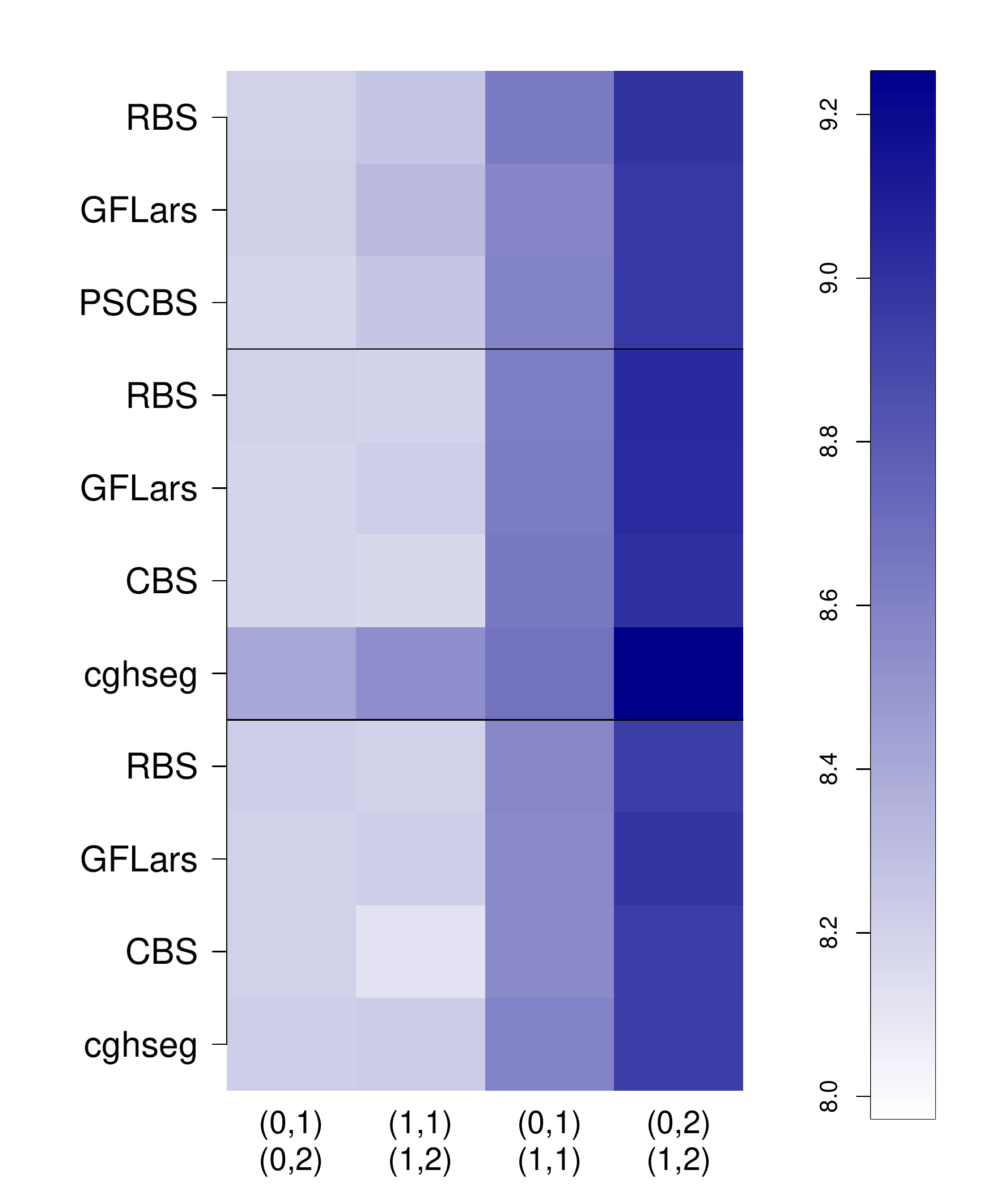}
\caption{$\log$(SNR) for missed (left) and caught (right) breakpoints for four types of breakpoints on data set 2 with 50\% normal cell contamination.}
\label{fig:SNR-50pct-Illu}
\end{figure}

An important question when using a biostatistical or bioinformatic tool is to assess its ability to recover events and to know which events they are likely to find and which of them are harder to detect.
In Table \ref{table:best-affy-illu} it can be seen that the pAUC is never at 100\%. This is not necessarily surprising as the signal is quite noisy and in fact considering noise level the pAUC is quite high. 
Figure \ref{fig:SNR-50pct-Illu} demonstrates that (as could be expected) missed change-points are those for which we have a low signal to noise ratio (the right panel is darker than the left panel).  
However, the signal to noise ratio substantially depends on the type of change-point.
Typically, in Figure \ref{fig:SNR-50pct-Illu} the column corresponding to the (0,2)-(1,2) transition is much darker than that of the (1,1)-(1,2) transition. This is confirmed by Table \ref{tab:prop-missed-bkp-2}, which indicates that for a high level of normal contamination in data set 2, the proportion of missed (1,1)-(1,2) change-points is greater than 1/2.

\begin{table}[ht]
\centering
\begin{tabular}{clcccc}
  \hline
 Statistic &Method & (0,1)-(0,2) & (1,1)-(1,2) & (0,1)-(1,1) & (0,2)-(1,2) \\ 
  \hline
  \multirow{3}{*}{$(c,d)$} & RBS  & \textcolor{red}{\bf{0.40}} & \textcolor{red}{\bf{0.47}} & \textcolor{red}{\bf{0.32}} & 0.31 \\ 
  & GFLars & 0.51 & 0.66 & 0.44 & 0.34 \\ 
  & PSCBS & 0.55 & 0.63 & 0.51 & 0.47 \\ 
  \hline
  \multirow{4}{*}{$(c)$} & RBS & 0.57 & 0.69 & 0.52 & 0.63 \\ 
  & GFLars  & 0.54 & 0.70 & 0.45 & 0.58 \\ 
  & CBS & 0.59 & 0.71 & 0.52 & 0.62 \\ 
  & cghseg  & 0.66 & 0.79 & 0.55 & 0.69 \\ 
  \hline
  \multirow{4}{*}{$(d)$} & RBS  & 0.49 & 0.54 & 0.39 & 0.24 \\ 
  & GFLars  & 0.49 & 0.51 & 0.34 & \textcolor{red}{\bf{0.20}} \\ 
  & CBS & 0.51 & 0.49 & 0.41 & 0.23\\ 
  & cghseg & 0.51 & 0.51 & 0.38  & 0.23 \\ 
   \hline
\end{tabular}
\caption{Proportion of missed breakpoints by method, statistic and type of copy-number transition (data set 2\add{, 50\% of tumor cells}).}
\label{tab:prop-missed-bkp-2}
\end{table}

\section{Summary and discussion}
\label{sec:summary-discussion}

We have developed a framework to assess the performance of various DNA copy number segmentation methods. A critical aspect of this framework is that it generates realistic copy-number profiles by resampling real SNP array data.  This allows \add{us} to study a large number of scenarios without relying on a particular statistical model. It is our opinion that this framework is simple to use as it depends on few parameters, all of which have a straightforward biological interpretation.  An {\code{R}} package is available and we believe that our proposed data generation scheme can be used readily as well as applied to other data sets and technologies. \add{It is also possible to extend the set of segmentation methods compared, as explained in the package documentation.} In this paper, we illustrated the usage of this framework on two \add{SNP array} data sets \add{from Affymetrix and Illumina}.

We were able to identify which technological and biological parameters drive the performance of segmentation methods.  First, it appears that the percentage of tumor cells in the sample plays a critical role: for a percentage lower than $70\%$, it is probably hopeless to recover the whole set of breakpoints with a high \suppress{precision}\add{accuracy}. We emphasize the relevance of the considered range of cellularity for applications: we expect that tumor cell lines should be well represented by the  100\% setting, while the 50\% is not unusual for clinical practice.
Second, it seems that different microarray technologies might lead to different performances.  Specifically, the ratio between the number of informative allelic probes \add{(heterozygous SNPs)} to the total number of probes is a crucial aspect, particularly for a high level of normal contamination. 
Finally, not all methods achieve similar performance across the scenarios that we have considered.  Interestingly, we show that methods that take advantage of both signal dimensions are generally but not always better than those using only one of them.  This variability between \add{segmentation} methods may be attributed to some extent to the biological and technological contexts, in the sense that some methods might be more adapted to certain scenarios. 

Our framework provides a way to critically evaluate the performance of segmentation methods, and therefore to rationally select one or several of them for a particular data set.  Such a quantitative assessment is also useful for interpretation.  For example, we showed that even in favorable scenarios, performances are not perfect.  Furthermore, perhaps unexpectedly, we showed that copy number transitions involving the gain or loss of a single DNA copy are not equally easy to recover, meaning that the proportion of different types of copy number transitions recovered by a particular segmentation method may not be directly interpretable.


\section{Acknowledgements}

The authors would like to thank Henrik Bengtsson and Cyril Dalmasso for very instructive discussions and feedback.
The authors are also grateful to the three referees whose constructive comments helped to improve the clarity of the paper. 

\clearpage

\bibliographystyle{BriefBiBib}
\bibliography{jointSeg}

\appendix
\section*{Appendix}
\addcontentsline{toc}{section}{Appendix}
\section{SNP array data sets}
\label{sec:snp-array-data}

\subsection{Data set 1}
\label{sec:affymetrix-data}
We have worked with a lung cancer data \cite{rasmussen11allele}, for which raw data is accessible at NCBI GEO database \cite{edgar02gene}, accession \href{http://www.ncbi.nlm.nih.gov/geo/query/acc.cgi?acc=GSE29172}{GSE29172}. DNA from patient-matched lung cancer and blood cell lines {\code{NCI-H1395}} and {\code{NCI-BL1395}} were mixed to simulate tumor tissue with 30, 50, 70, 100\% cancer cells. DNA was analyzed on Affymetrix SNP6.0 microarray. Data were normalized using  ASCRMAv2 \cite{bengtsson09a-single-array} followed by TumorBoost \cite{bengtsson10tumorboost}.
For the sake of reproducibility, the {\code{R}} scripts that were written to normalize this data set are distributed in the {\code{jointseg}} package, together with the normalized data itself. 

\subsection{Data set 2}
\label{sec:illumina-data}
We have also worked with a breast cancer data \cite{staaf08segmentation},  for which raw data is accessible at NCBI GEO database \cite{edgar02gene}, accession \href{http://www.ncbi.nlm.nih.gov/geo/query/acc.cgi?acc=GSE11976}{GSE11976}. DNA from patient-match breast cancer cell line ({\code{HCC1395}}) and its match normal  {\code{HCC1395BL}} were mixed to simulate tumor tissue with 14, 34, 50, 79, 100\% cancer cells.  DNA was analyzed on Illumina HumanCNV370-Duov1 microarrays. We obtained the BAF-normalized and summarized data as calculated by the Illumina BeadStudio software~\cite{illumina-inc06illuminas,illumina-inc07genotyping,peiffer06high-resolution}

\subsection{Description of annotated copy-number regions}
The list below describes the different copy number states available for data generation. They are labeled as a pair $(c_1,c_2)$, where $c_1$ corresponds to the minor copy number (the smallest of the two parental copy numbers), and $c_2$ corresponds to the major copy number (the largest of the two) \cite{neuvial11statistical}. 
\begin{description}
\item[(1,1):] normal (one copy from each parent)
\item[(0,1):] hemizygous deletion (loss of one parental copy)
\item[(0,0):] homozygous deletion (loss of both parental copies)
\item[(0,2):] copy-neutral LOH (loss of one parental copy and gain of the other)
\item[(0,3):] loss of one parental copy and gain of two copies from the other parent)
\item[(1,2):] single copy gain
\item[(1,3):] unbalanced two-copy gain (gain of two copies from the same parent)
\item[(2,2):] balanced two-copy gain (gain of one copy from each parent)
\item[(2,3):] three-copy gain (gain of one copy from each parent, and two copies from the other parent)
\end{description}
~
\begin{table}[!htp]
\centering\small
\begin{tabular}{rrrrrrrrrr}
  \hline
  CN state & (0,1) & (0,2) & (0,3) & (1,1) & (1,2) & (1,3) & (2,2) & (2,3) & (0,0) \\ 
  \hline
  Data set 1 & 22615 & 24135 & 25405 & 21539 & 19048 & 20903 & 27924 & 31098 &   0 \\ 
  Data set 2 & 2492 & 5484 & 6545 & 3196 & 2746 &   0 & 3044 &   0 & 838 \\ 
   \hline
\end{tabular}
\caption{Size of annotated copy-number regions for each of the 2 data sets.} 
\label{tab:regData}
\end{table}

\section{Reproducing the figures and tables of this paper}
\begin{verbatim}
library(jointseg)
path <- system.file("figures", package="jointseg")
filenames <- list.files(path, pattern="*.R$")
for (filename in filenames) {
  print(filename)
  pathname <- file.path(path, filename)
  source(pathname, local=TRUE)
}
\end{verbatim}

\add{The scripts used for the performance evaluation reported in this paper are available in the subdirectory \code{"eval"} of the \code{jointseg} package:}

\begin{verbatim}
path <- system.file("eval", package="jointseg")
\end{verbatim}

\section{Session information}
\begin{verbatim}
> sessionInfo()
R version 3.0.2 (2013-09-25)
Platform: x86_64-apple-darwin10.8.0 (64-bi)

locale:
[1] fr_FR.UTF-8/fr_FR.UTF-8/fr_FR.UTF-8/C/fr_FR.UTF-8/fr_FR.UTF-8

attached base packages:
[1] parallel  stats     graphics  grDevices utils     datasets  methods
[8] base

other attached packages:
 [1] PSCBS_0.39.1       DNAcopy_1.36.0     PSCN_1.0.1         MASS_7.3-29
 [5] changepoint_1.1    zoo_1.7-10         cghseg_1.0.1       jointseg_0.5.1
 [9] acnr_0.1.4         R.utils_1.27.5     R.oo_1.15.8        R.methodsS3_1.5.2
[13] matrixStats_0.8.12

loaded via a namespace (and not attached):
[1] grid_3.0.2      lattice_0.20-24 R.cache_0.9.0
\end{verbatim}

\end{document}